\documentclass[aip,rsi,amsmath,amssymb,reprint]{revtex4-1}

\usepackage{graphicx}
\usepackage{dcolumn}
\usepackage{bm}
\usepackage{subcaption}
\usepackage{float}
\usepackage{wrapfig}

\begin{document}

\title{Identify Statistical Similarities and Differences Between the Deadliest Cancer Types Through Gene Expression}

\author{Arturo Chavez}
 \affiliation{Department of Electrical Engineering and Computer Science, Massachusetts Institute of Technology}
 
\author{Dimitris Koutentakis}
 \affiliation{Department of Electrical Engineering and Computer Science, Massachusetts Institute of Technology}

\author{Youzhi Liang}
 \affiliation{Department of Mechanical Engineering, Massachusetts Institute of Technology}
 
\author{Sonali Tripathy}
 \affiliation{Sloan School of Management, Massachusetts Institute of Technology}
 
\author{Jie Yun}
 \affiliation{Department of Civil and Environmental Engineering, Massachusetts Institute of Technology}

\date{\today}

\begin{abstract}
Prognostic genes have been well studied within each type of cancer. However, investigations of the similarities and differences across cancer types are rare. In view of the optimal course of treatment, classification of cancers into subtypes is critical to the diagnosis. We examined the properties in gene co-expression networks using a patient-to-patient correlation network analysis and a weighted gene correlation network analysis (WGCNA) for five cancer types using data generated by UC Irvine. We further analyze and compare the degree, centrality and betweenness of the network for each cancer type and apply a multinomial logistic regression to identify the critical subset of genes. Given the cancer types provided, our study presents a view of emergent similarities and differences across cancer types. 
\end{abstract}

\maketitle

\section{Introduction}
Cancer describes a collection of diseases that share some common characteristics, particularly unregulated cell growth,  they also vary widely in terms of mortality rate, treatment options, and prevalence in the population. Accurate diagnosis of cancer type is essential to decide treatment options, therapy and prognoses. However, some cancers are difficult to distinguish based on a single test \cite{khan2001classification}. Gene expression microarray technology provides precise information for cancer prognosis and treatment, and has been used to categorize cancers into subgroups\cite{lee2003classification}. Current classification methods include nearest prototype classifier by defining subset of genes that best characterize each class \cite{tibshirani2002diagnosis}, supervised classification algorithms to identify gene expression signature, and the use of combined algorithms \cite{dubey2015breast}. 

These methods have experienced moderate success, so clearly the methods are identifying relevant statistical differences in the tumor types in order to classify them correctly. Digging one level deeper, we are interested to explore the statistical differences between tumors, tying them to phenotype differences in disease outcomes. Using this data-driven approach, we aim to understand the variation within tumors of the same type as well as the consistent differentiating features that distinguish each tumor type. 

\subsection{Tumor Classification Background}  
Khan et al. \cite{khan2001classification} explore the use of neural network classification models to classify cancer subtypes taking cDNA expression data as the input. Their analysis was specific to small blue-cell tumors (SBCTs) which can be further classified into neuroblastoma (NB), rhabdomyosarcoma (RMS), non-Hodgkin lymphoma (NHL) and the Ewing family of tumors (EWS). Correct classification into subtypes is critical to selecting the optimal course of treatment. Usual methods for tumor classification often use spectroscopy, but SBCTs are challenging to classify visually. There have been many attempts to use gene-expression data to aid in classification, but so far none have been proven to be effective in identifying cancers that belong to several categories.

They started with a panel of 6567 genes from which to find meaningful features. In order to make the dimensionality of the data more manageable, they eliminated genes that had expression levels below a threshold. With the remaining 2308 genes, they performed PCA to further reduce the dimensionality, taking the largest 10 components which accounted for 63\% of the variation. After training on these features, their model was able to fit all of the 63 samples from their training set. To identify the most important genes, the authors altered each of the locations to measure the overall classification’s sensitivity to that gene. After identifying the most important genes, the authors performed multidimensional scaling (MDS) to visualize the clear separation between cancers. 

When they tested the model’s ability to classify new samples, they were pleased to be able to classify all the cancer types correctly. Unfortunately, they were unable to reach the level of 95\% confidence level in the diagnosis that they were targeting. This highlights the challenge of using machine learning methods in the medical field, since clinical use needs highly reliable AI systems. This study motivates further work of this kind with other disease types and larger data sets. Also, this result of a reasonably successful classification method motivates our analysis of the statistical properties of the different tumor gene expression profiles, from which these classifiers form decision boundaries.

\subsection{Network Methods Background}

Specifying features in genetics is a challenge because there are often complicated interactions between genes. To understand these relationships researchers have used network models. Juan A. Botia et al.\cite{botia2018g2p} analyzed 1126 genes relating to 25 subtypes of Mendelian neurological disease defined by Genomics England (March 2017) together with 154 gene-specific features capturing genetic variation, gene structure and tissue-specific expression and co-expression. He developed a technique to identify the gene mutations that can lead to a neurological disorders. Random samples were selected with no disease association to develop decision tree models for each subtype. Within the disorder subtypes, network models were used to improve the predictive power.

Another instance of network approaches in genomes, Yang et al. \cite{yang2014gene} applied the weighted correlation network analysis (WGCNA) method to construct a gene co-expression network. In this study, they primarily investigated the prognostic genes that distinguish between cancers. They investigated these genes with three distinct levels of depth properties: specific genes, gene modules, and the system holistically. At the gene level, they found that network properties could distinguish prognostic genes from other genes. More specifically, using Fisher's exact test, they were able to conclude that prognostic genes tend not be hubs in the co-expression network. On the gene modules level, they discovered that prognostic genes are enriched significantly. Third, on the system level, some prognostic modules are conserved across tumour types.

\section{Preliminary Methods}
\subsection{Dataset Description}
The dataset is provided by University of California at Irvine and is located \href{https://archive.ics.uci.edu/ml/datasets/gene+expression+cancer+RNA-Seq#}{here}. 
The data includes 801 samples, each with 20,532 gene positions. Each sample vector contains the RNA-Seq gene expression levels. Each sample in the dataset corresponds to a particular tumor type. Every sample is one of five types: breast invasive carcinoma (BRCA), kidney renal clear cell carcinoma (KIRC), colon adenocarcinoma (COAD), lung adenocarcinoma (LUAD), and prostate adenocarcinoma (PRAD).

\subsection{Understanding the data}

\begin{figure}[H]
\vspace{-5mm}
\centering
\begin{subfigure}{0.48\linewidth}
    \centering
    \includegraphics[width = \textwidth]{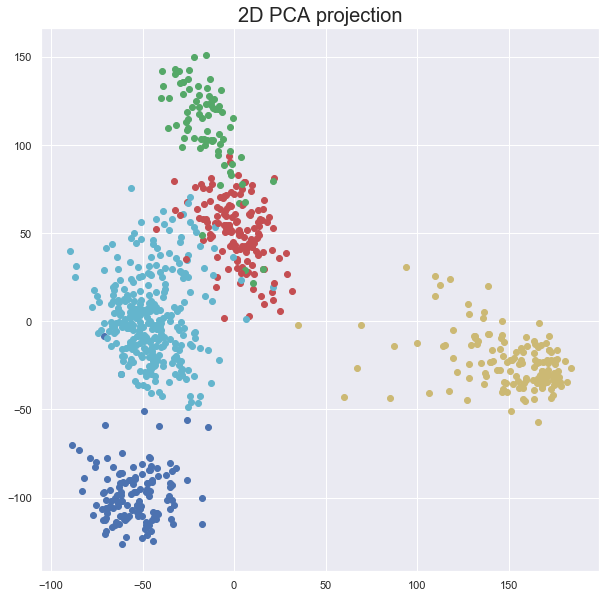}
    \caption{2D PCA data projection}
    \label{fig:2dPCA}
\end{subfigure}
\begin{subfigure}{0.48\linewidth}
    \centering
    \includegraphics[width = 0.96\textwidth]{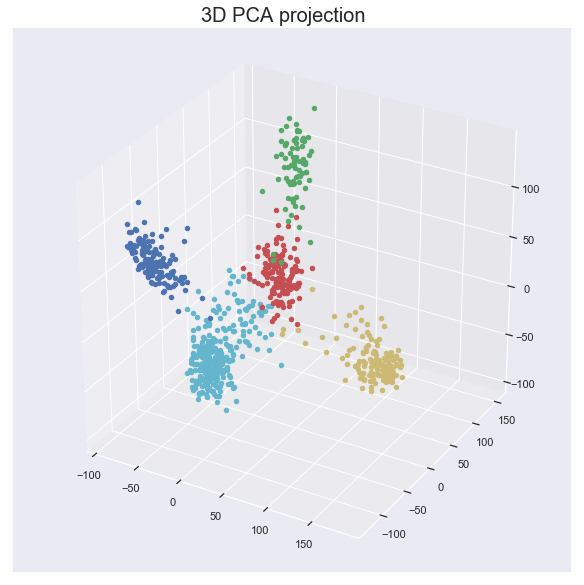}
    \caption{3D PCA data projection}
    \label{fig:3dPCA}
\end{subfigure}
\vspace{-4mm}
\label{fig:PCAs}
\end{figure}

A preliminary analysis of our data is shown in Fig.~\ref{fig:2dPCA}. Not unusual in the genomics setting, we run into the curse of dimensionality, making our $801 \times 20,532$ dimensional matrix difficult to visualize. We performed Principal Component Analysis on our data set and plotted the projection of our data on the 2  principal components with the largest corresponding eigenvalues in Figure \ref{fig:2dPCA}, and  the projection of our data on the 3  principal components with the larges eigenvalues in Figure \ref{fig:3dPCA}. Prior to performing the eigenvector decomposition, we pre-process our data by subtracting the column mean from the each entry. The result is the matrix X with dimension $801 \times 20,532$ with each column having mean 0. Taking the eigenvector decomposition we get $X = V \Lambda  V^T,$ where $\Lambda$ is a diagonal matrix of the eigenvalues (sorted such that the largest eigenvalue is in the top left) and $V$ has the corresponding eigenvectors as its columns. Taking the first $d$ columns of V, we get $T_d = X V_d$, where $T$ has dimension $n \times d.$ 

We compute $T_2$ and $T_3$ and plot the results. In both of the plots, each point was colored according to what type of cancer it represents. This further validates our intuitions that the each cancer type has particular features that distinguish it from the others.

\subsection{Variance of Tumor Types in Reduced Dimension}
We aim to understand how the various tumor types differ, both in statistical and phenotypic terms. The previous PCA results show that projecting the samples onto the first two or three principal components lead to a reasonably clean separation. Interestingly, some cancer types appear to be clustered more tightly together in this lower dimensional space, while others appear to be more loosely dispersed. Also, it is interesting to note which pairs of tumor types appear closer together in this space. To quantify both of these notions, we fit a Gaussian mixture model to the PCA-transformed points. Using $T_2$ from the previous section, we fit a five Gaussian mixture that appears to closely approximate the true labeling of the points. Sampling from a Gaussian mixture can be thought of as a two step process. First, it involves sampling from a multinomial distribution with parameters $\pi$ (similar to an unfair dice). The result of the first step determines which Gaussian to sample from in the second step. Therefore, the conditional probability of the coordinates of a sample, given it is a particular cancer type $c$, is distributed according to $\mathcal{N}(\mu_c, \sigma_c^2)$. The second step is simply to sample from that Gaussian. Gaussian mixture models are fit using the expectation-maximization (EM) algorithm, where the objective is to maximize the log-likelihood of generating the training data. Fitting this model results in the parameters $\pi, \mu, \Sigma$ for each type of cancer.

\begin{figure}[H]
    \centering
    \includegraphics[width=\linewidth]{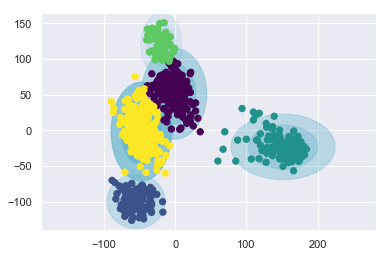}
    \caption{Above is the visualization of the gaussian mixture model fit to explain the distribution of the samples in the 2 dimensional space specified by the first principal components. PRAD is blue, LUAD is purple, BRCA is yellow, KIRC is teal, and COAD is green.}
    \label{fig:Gaussian}
\end{figure}

Assuming that this fit is reasonable, we can quantify the notions of homogeneity within a tumor type by inspecting the covariance matrix of the Gaussian corresponding to that cluster of samples.  Because we are interested in the variance along the axis of the principal components, we constrain the Gaussians to be oriented along those axes (forcing the covariance matrices to be diagonal). We get the following results where the vector is ordered [LUAD, PRAD, KIRC, COAD, BRCA]:
$$ \pi^T = \begin{pmatrix} 0.23 & 0.16 & 0.18 & 0.08 & 0.35
 \end{pmatrix} $$
$$\mu_{L} = \begin{pmatrix} 
  -1.60 \\  51.24
\end{pmatrix}, \mu_{P}\begin{pmatrix} 
  -54.91 \\ -100.11 
\end{pmatrix}, \mu_{K} = \begin{pmatrix} 
  151.45 \\ -23.29 
\end{pmatrix}$$ 
$$ \mu_{C} = \begin{pmatrix} 
  -19.50 \\ 120.05 
\end{pmatrix}, \mu_{B} = \begin{pmatrix} 
  -47.47 \\ -1.81
\end{pmatrix}$$
$$ \Sigma_{L} = \mbox{Diag} \binom{233.83} {451.96}, \Sigma_{P} = \mbox{Diag} \binom{ 186.51}{ 160.41},$$ 
$$ \Sigma_{K} = \mbox{Diag}\binom{594.35}{235.63}, \Sigma_{C} = \mbox{Diag} \binom{92.20}{ 224.27}, $$
$$ \Sigma_{B} = \mbox{Diag} \binom{201.54}{536.06}. $$
The model was able to fit the data reasonably well, which can be seen in Figure \ref{fig:Gaussian}. For example, the weights $\pi$ are nearly the same distribution as the true labels which are: $ \begin{pmatrix}  0.18 & 0.17 & 0.18 & 0.10 & 0.37 \end{pmatrix}.$ Interestingly, we see that KIRC has the largest variance in the first principal component and BRCA has the largest variance in the second principal component. Overall it appears that KIRC is the most dispersed cancer type, since it has largest variance on average over the 2-D space. This can be evidenced in Figure \ref{fig:Gaussian}, seeing that there are several samples that are more than 2 standard deviations away from the cluster mean. COAD and PRAD are the most concentrated, suggesting that those samples are more homogeneous. 

Note, that our model is not a perfect fit. Comparing our fitted GMM model to the 2-D projection of the data in Figure \ref{fig:2dPCA}, we see that some COAD samples are found within the LUAD cluster. This GMM model does not explain this feature of the data, motivating the use of other methods to understand the gene expression profiles of these cancer types.

\subsection{Stochastic Neighbor Embedding}
In addition to PCA, we perform t-distributed stochastic neighbor embedding to our data. tSNE is a probabilistic approach to place objects from high-dimensional space into low-dimensional space so as to preserve the identity of the neighbors. Prior to tSNE, stochastic neighbor embedding (SNE) was proposed, which used the same general approach by placed a Gaussian on each object in high-dimensional space. This resulted in the ``crowding problem" where many points would be mapped together in the center. To overcome this problem, Hinton et al.\cite{maaten2008visualizing} proposed tSNE which has larger tails and a steeper drop moving away from the mean (within close range). Both methods are fit by minimizing the KL divergence between the low and high dimensional probabilities of picking a particular neighbor. Intuitively, this method keeps ``nearby" points in high dimension close to each other in low dimensional space, while keeping separated points relatively far apart in the low dimensional space. In this case tSNE is able to separate the tumor types with high precision (notably better than PCA). This result supports our intuition that different cancer types are statistically distinguishable. For the rest of the paper, we aim to characterize those statistical differences more precisely. 

\begin{figure}[H]
    \centering
    \includegraphics[scale=.4]{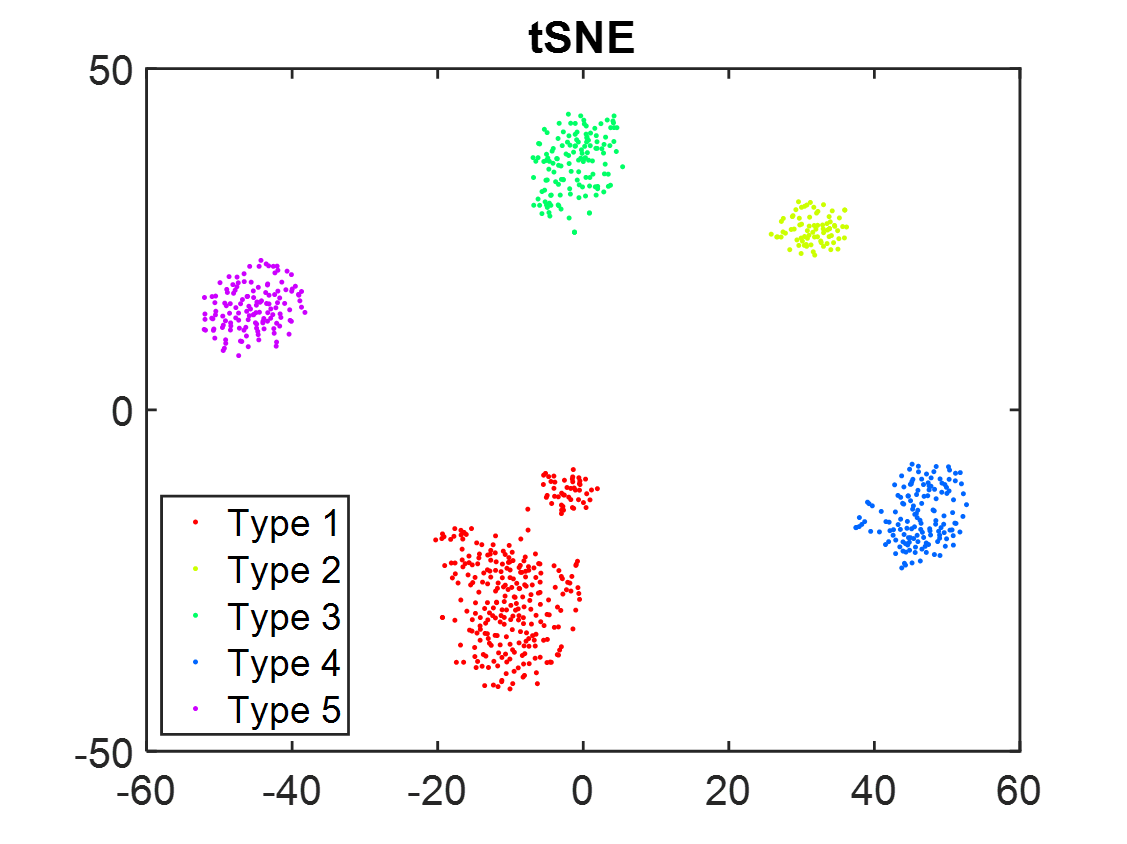}
    \caption{tSNE on our data gives the following well-separated clusters}
    \label{fig:tSNE}
\end{figure}

\subsection{Hierarchical Clustering}

\begin{figure}[H]
    \centering
    \includegraphics[scale=0.5]{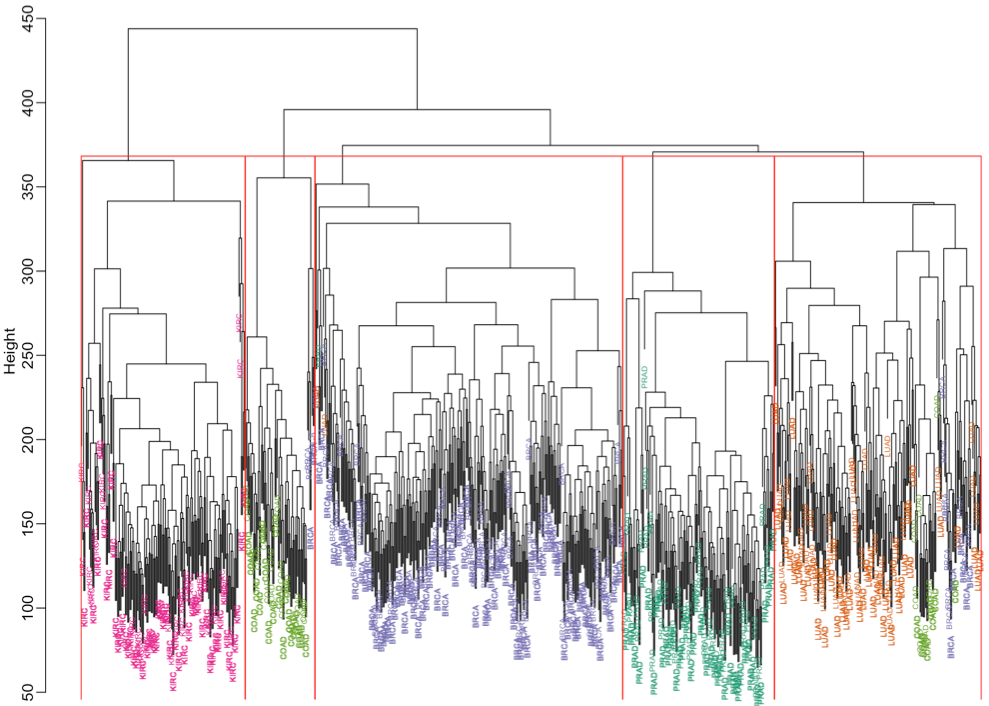}
    \caption{Hierarchical clustering of the gene data set where pink is KIRC, green is COAD, purple is BRCA, teal is PRAD, and orange is LUAD.}
    \label{fig:Hierarchical Clustering Dendrogram}
\end{figure}

Hierarchical clustering of gene expression is a popular mechanism to cluster genes with similar expression patterns together. This clustering mechanism involves calculation of distance between two gene vectors to find the similarity between them. The dendrogram was sliced at a height of 370 to find five clusters in particular. Figure \ref{fig:Hierarchical Clustering Dendrogram} demonstrates the samples clustered into five clusters where each cancer type is majorly clustered into just one cluster. There seems to be one of the clusters that consists of more than one cancer type, signifying that in some patients the distance between the gene vectors is close enough. These cancer types are BRCA, LUAD and COAD. From figure \ref{fig:Gaussian} also it could be seen that these three cancer types are close to each other.

\section{Network Analysis}

\subsection{Patient-to-Patient Correlation Network} 
\begin{figure}[H]
\centering
\begin{subfigure}{0.4\linewidth}
    \includegraphics[width=\textwidth]{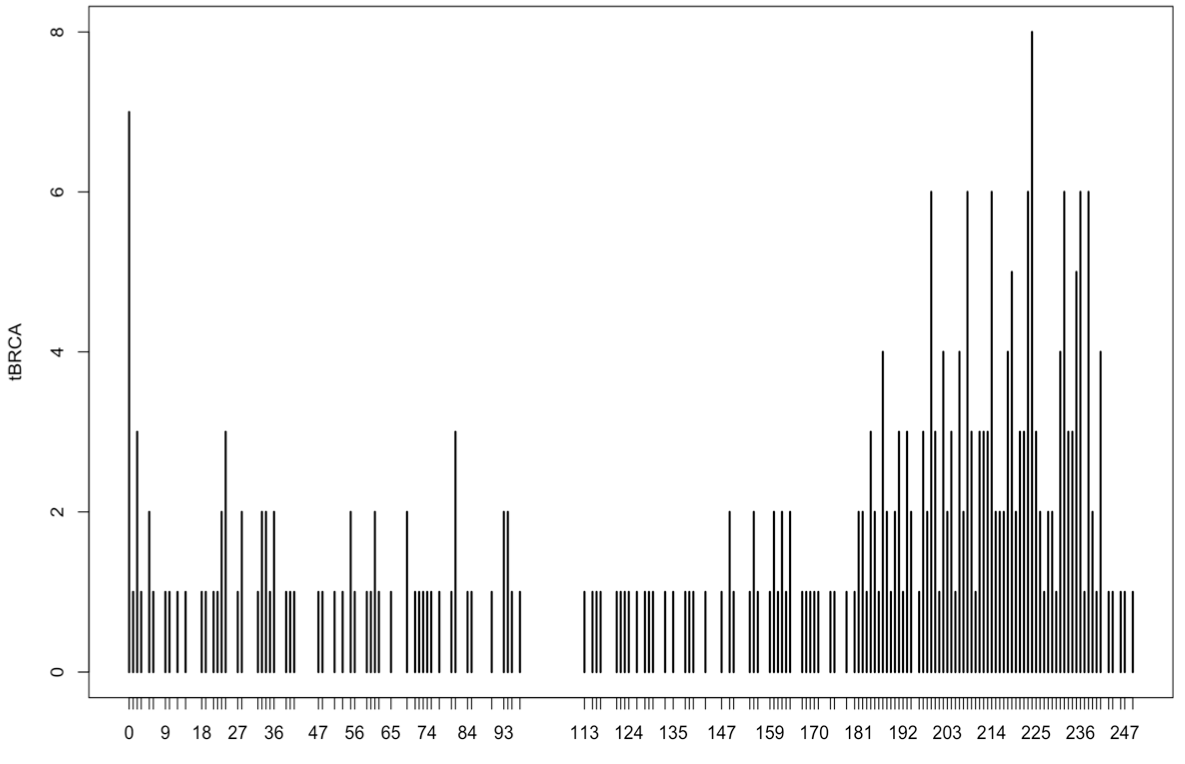}
    \caption{BRCA}
    \label{fig:brcadeg}
\end{subfigure}
\begin{subfigure}{0.4\linewidth}
    \includegraphics[width=\textwidth]{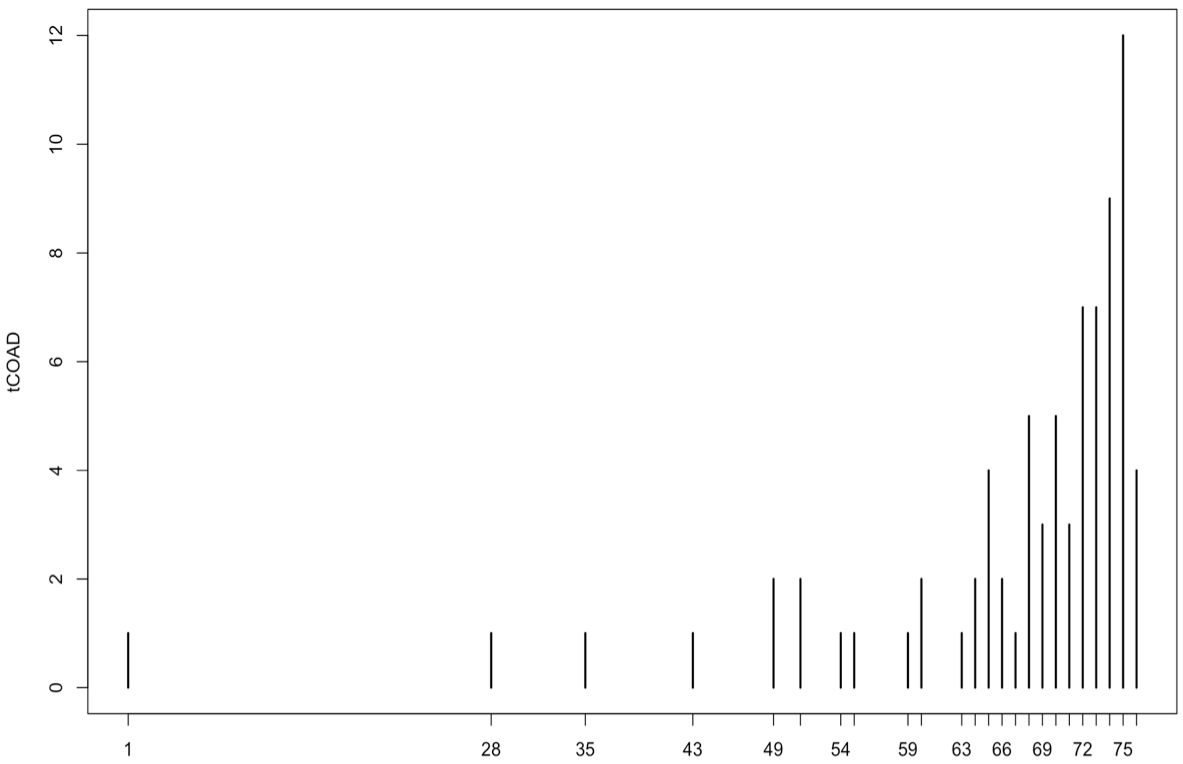}
    \caption{KIRC}
    \label{fig:kircdeg}
\end{subfigure}

\begin{subfigure}{0.3\linewidth}
    \includegraphics[width=\textwidth]{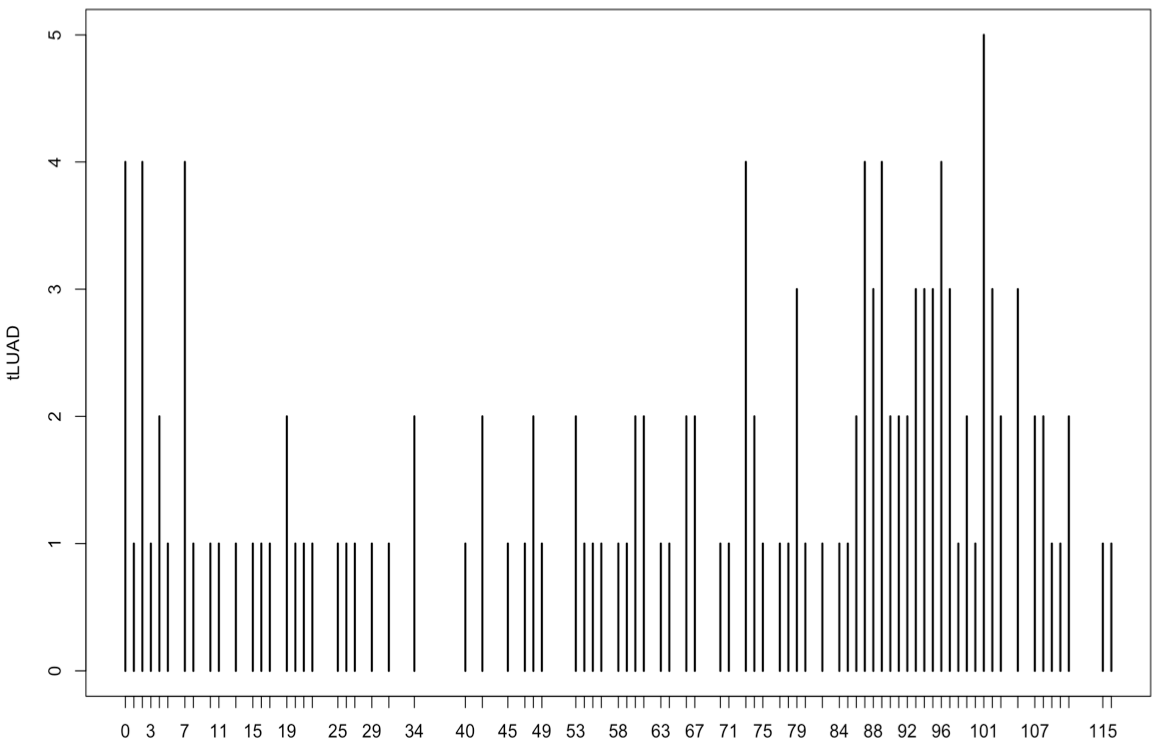}
    \caption{LUAD}
    \label{fig:luaddeg}
\end{subfigure}
\begin{subfigure}{0.3\linewidth}
    \includegraphics[width=\textwidth]{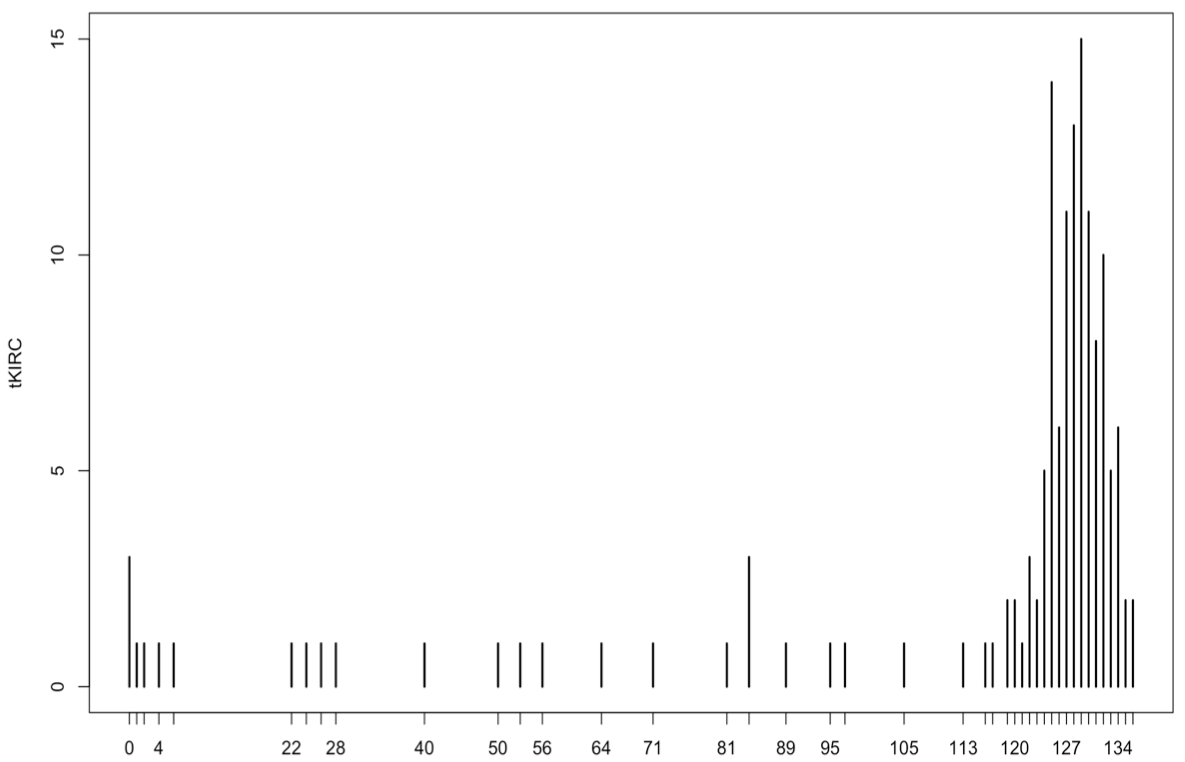}
    \caption{COAD}
    \label{fig:coaddeg}
\end{subfigure}
\begin{subfigure}{0.3\linewidth}
    \includegraphics[width=\textwidth]{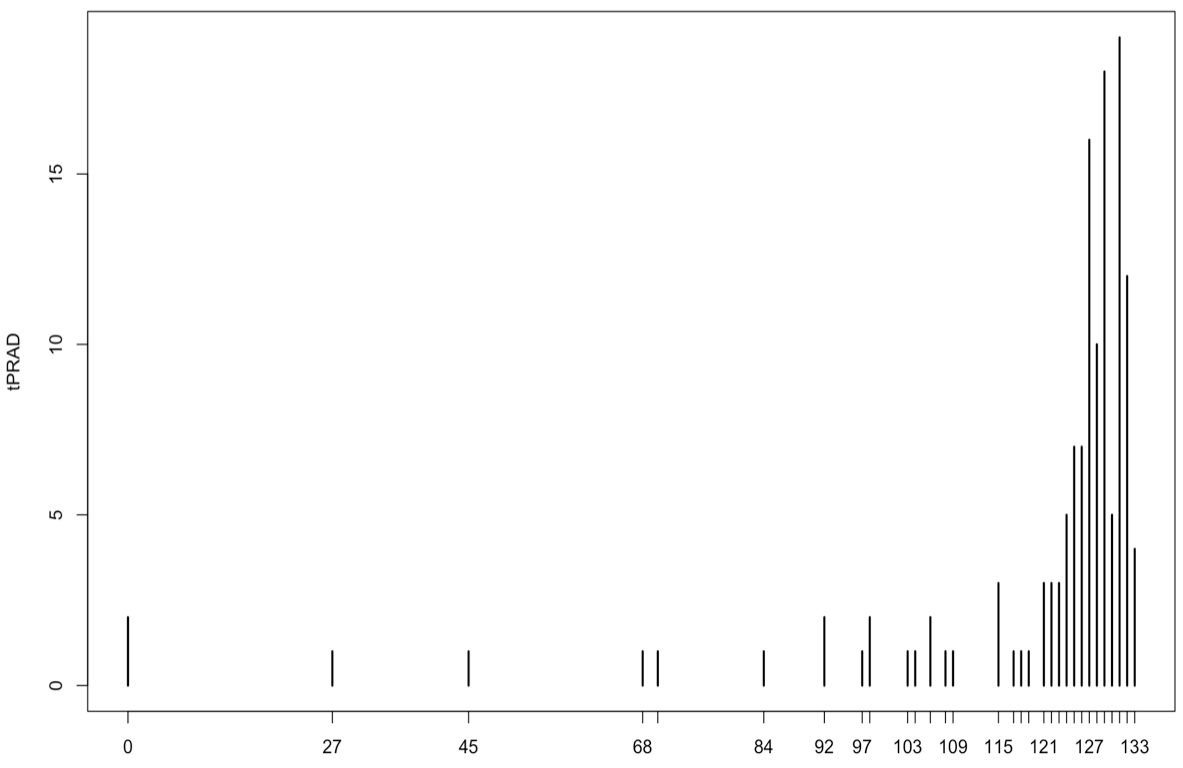}
    \caption{PRAD}
    \label{fig:praddeg}
\end{subfigure}
\caption{Degree distribution of network for each cancer type }
\label{fig:DegDist}
\vspace{-5mm}
\end{figure}

In order to understand the relationships between the samples in our data set, we constructed a network with each sample representing a node. The edges between $S$ samples are determined by the level of the correlation between the $G \times 1$ dimensional gene expression vectors. We start with our data matrix $A$ which is $S \times G$. Our correlation coefficients are defined as,
$$ \rho_{i,j} = \frac{ \sum^G_{k=1} (A_{i,k} - \mu_i)(A_{j,k} - \mu_j)}{\sigma_i \sigma_j}.$$
Given the correlation $\rho_{i,j}$ between gene expression vector for sample $i$ with the gene expression vector for sample $j$, we define a threshold value, drawing an edge between sample $i$ and sample $j$ if the correlation is statistically significant. We determine whether a correlation coefficient is significant using the Fisher transformation, which converts the distribution of Pearson's correlation coefficients to a normal distribution. This transformation takes the following form:
$$ Z_{i,j} = \frac{1}{2} \ln ( \frac{1+\rho_{i,j}}{1- \rho_{i,j}}) $$
Using the transformed correlation coefficients we can obtain a p-value from the Z-score, since they correspond to the normal distribution. We chose the 5\% significance level to draw our edges in this graph. 
Figure \ref{fig:DegDist} shows the degree distributions of the networks created by the mechanism discussed above for each cancer type. The degree distributions are left skewed suggesting that there are many high degree nodes among all the 801 patients. This also helps us conclude that the change in gene expression levels is highly similar for patients within one cancer type. Furthermore, the centrality measures are summarized in Table \ref{PatTable}. These measures suggest that there are certain patients that are central to the network corresponding to the cancer type, and thus are representative samples. We expect that adding more patients to the network would change the degree distributions and centralities of each patient. Then, depending on the degree measures of the new patient relative to patients close to this new patient who were already present in the network, we should be able to classify these patients into groups that would require similar therapies.

\begin{table}[H]
\begin{tabular}{|c|c|c|c|}
    \hline
Cancer & Degree & Eigenvector & Pagerank\\
    \hline
   PRAD & 34,158,275,390 & 34,158,275,390 & 34 \\
   BRCA & 99 & 111 & 99 \\
   LUAD & 229 & 229 & 229\\
   KIRC & 423,591 & 376,591 & 376\\
   COAD & 26,237,264,665 & 665 & 237,264,665\\
    \hline
\end{tabular}
\caption{Table summarizing nodes with max. Centralities}
\label{PatTable}
\end{table}

\subsection{Weighted Gene Co-expression Network Analysis}

A commonly used technique to analyze such data sets is to create a Weighted Gene Co-expression Network \cite{fuller2007weighted,langfelder2008wgcna}. This is a graph which has genes as nodes and the edge is given between two nodes represents the correlation between the two nodes that the edge joins. In order to build such a network, we start by first splitting our data set based on cancer type and then proceed with the correlation computation as described above.

\begin{figure}[H]
    \centering
    \begin{subfigure}{0.48\linewidth}
        \includegraphics[width=\textwidth]{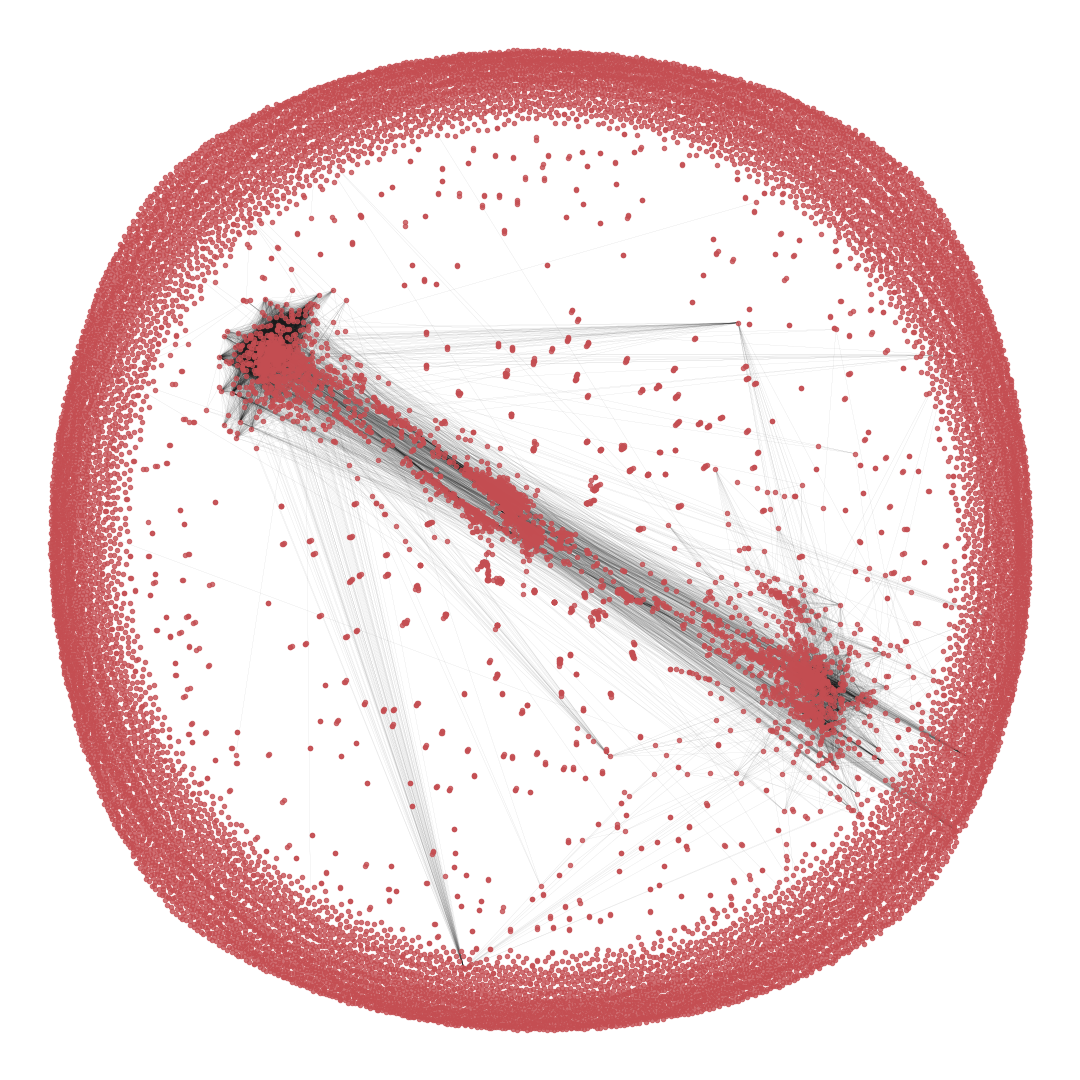}
        \caption{Breast cancer graph}
        \label{fig:brcanet}
    \end{subfigure}
    \begin{subfigure}{0.48\linewidth}
        \includegraphics[width=\textwidth]{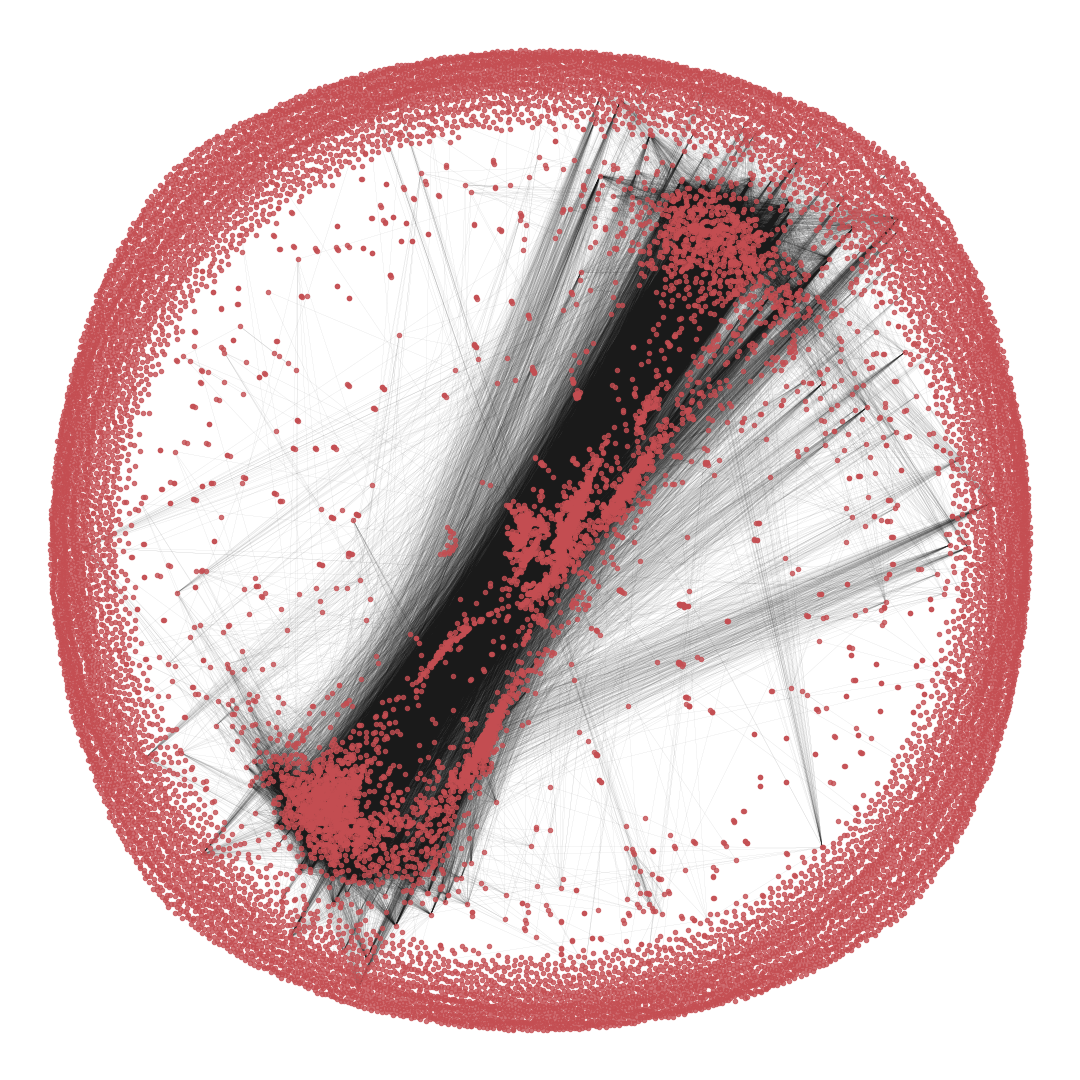}
        \caption{Kidney cancer graph}
        \label{fig:kircnet}
    \end{subfigure}
    
    \begin{subfigure}{0.3\linewidth}
        \includegraphics[width=\textwidth]{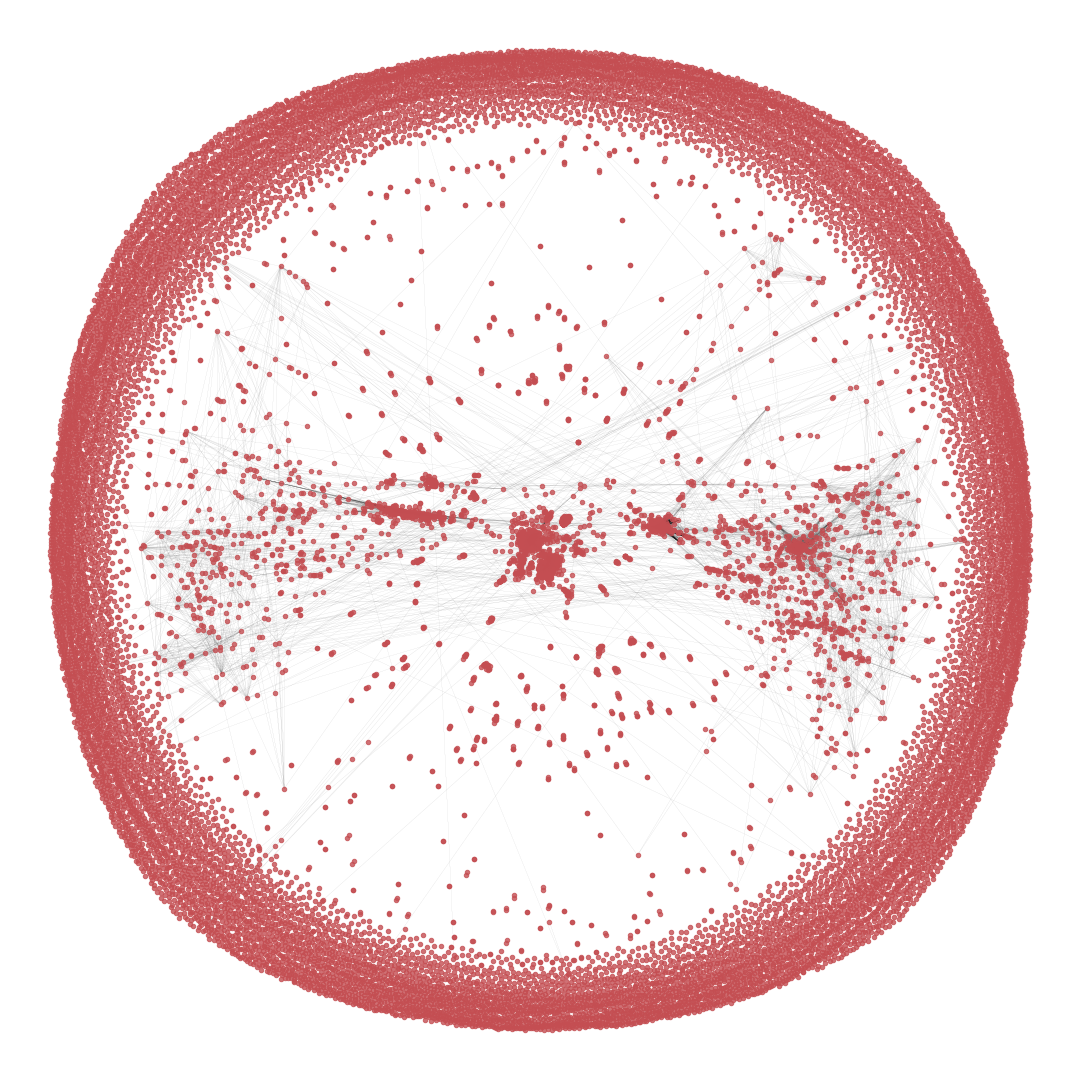}
        \caption{Lung cancer graph}
        \label{fig:luadnet}
    \end{subfigure}
    \begin{subfigure}{0.3\linewidth}
        \includegraphics[width=\textwidth]{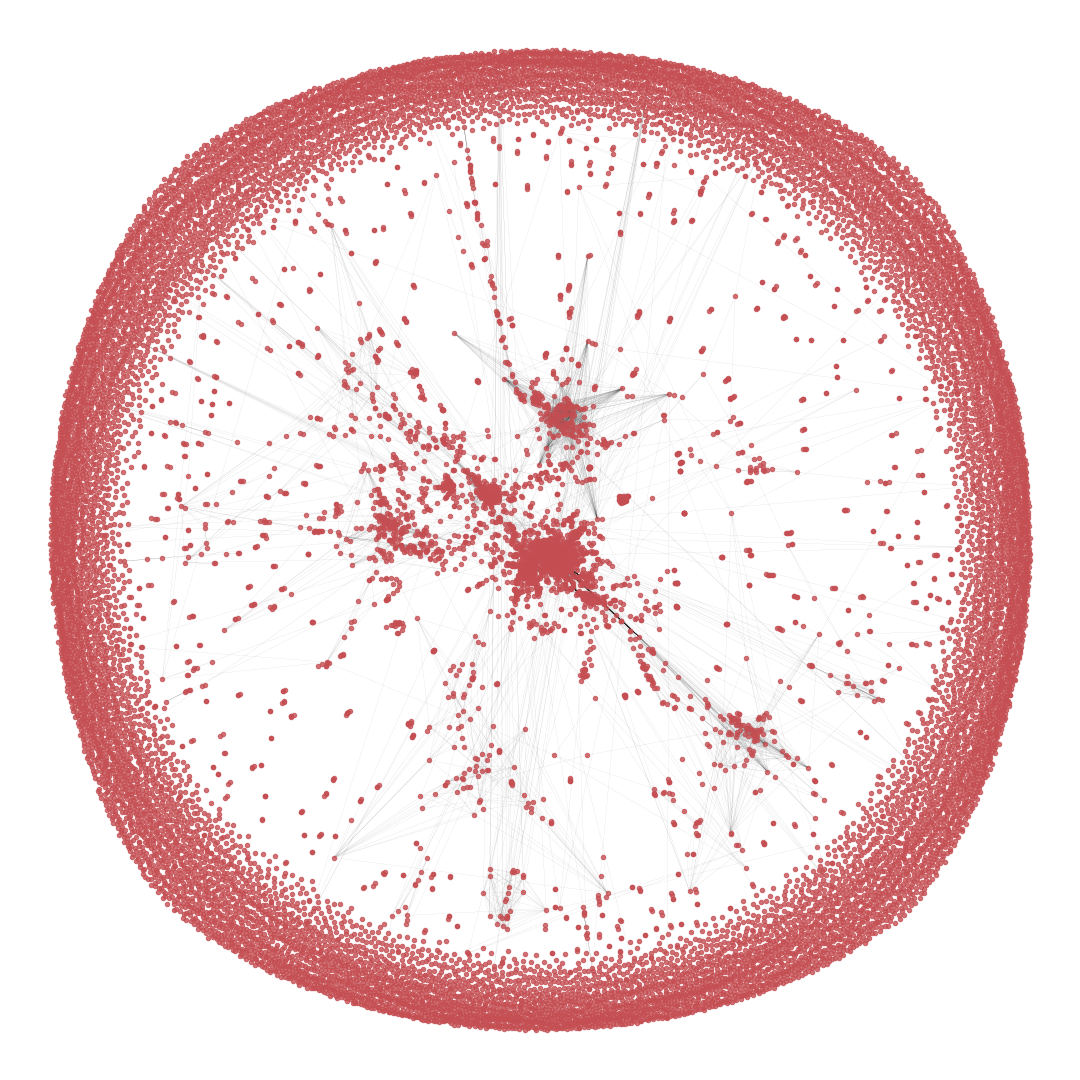}
        \caption{Colon cancer graph}
        \label{fig:coadnet}
    \end{subfigure}
    \begin{subfigure}{0.3\linewidth}
        \includegraphics[width=\textwidth]{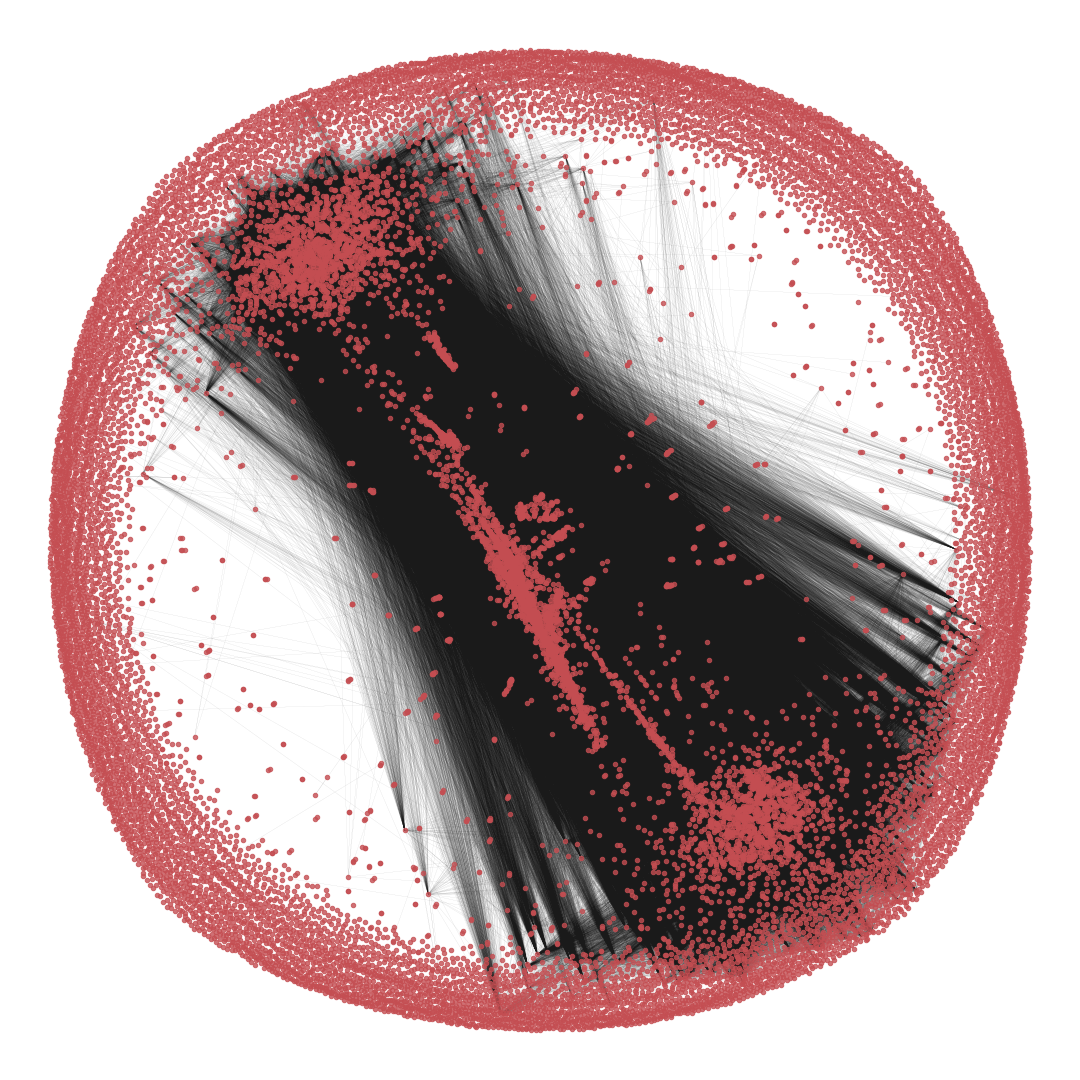}
        \caption{Prostate cancer graph}
        \label{fig:pradnet}
    \end{subfigure}
    \caption{WGCNs for each cancer type}
    \label{fig:WGCNA}
\end{figure}

For the matrix containing the gene expressions for each subset of our data, we compute the Pearson correlation matrix, as shown above and then use that as our preliminary adjacency matrix. Once we have computed the matrix, we build the network by adding all the nodes, but only draw edges if the correlation is above a value of $\rho_{X,Y}>0.8$. This threshold was chosen based on Fisher exact test leads to a significance level of around $5\%$. Additionally, it yields graphs that are sparse enough to visualize, though also dense enough that will allow us to make accurate computations. The networks resulting from this method are shown in Figure \ref{fig:WGCNA}. It is worth noting here that creating such large networks, plotting and computing the centrality measures proved to be very computationally intensive, as they all had above 20,000 nodes and between 40,000 and 200,000 edges. Even when using a powerful server (courtesy of the MIT Math Dep.), the algorithms took hours to run for each of the networks. 

It is interesting to note that even when only plotting the edges above the 0.8 correlation threshold, the graphs seem very dense. This is partially caused by the fact that some genes are naturally correlated and would be connected in the graph anyways. A way to go around this would be to use the partial correlation matrix as the adjacency matrix instead. The partial correlation would effectively condition on the rest of the genes, resulting in a more sparse network. However computing the partial correlation proved to be much harder computationally, or even impossible. 

\begin{figure}[H]
    \centering
    \begin{subfigure}{0.48\linewidth}
        \includegraphics[width=\textwidth]{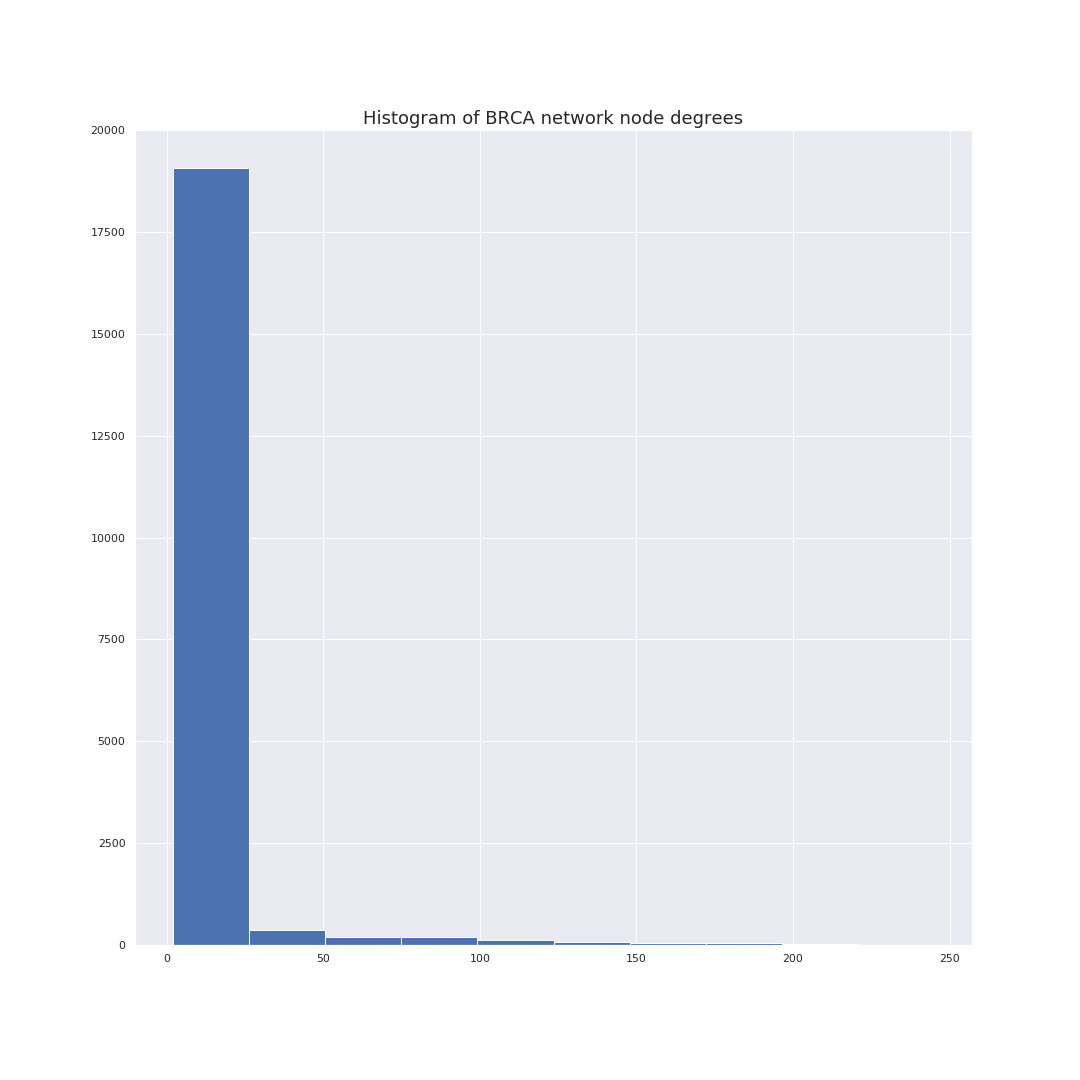}
        \caption{BRCA}
        \label{fig:brcadeg}
    \end{subfigure}
    \begin{subfigure}{0.48\linewidth}
        \includegraphics[width=\textwidth]{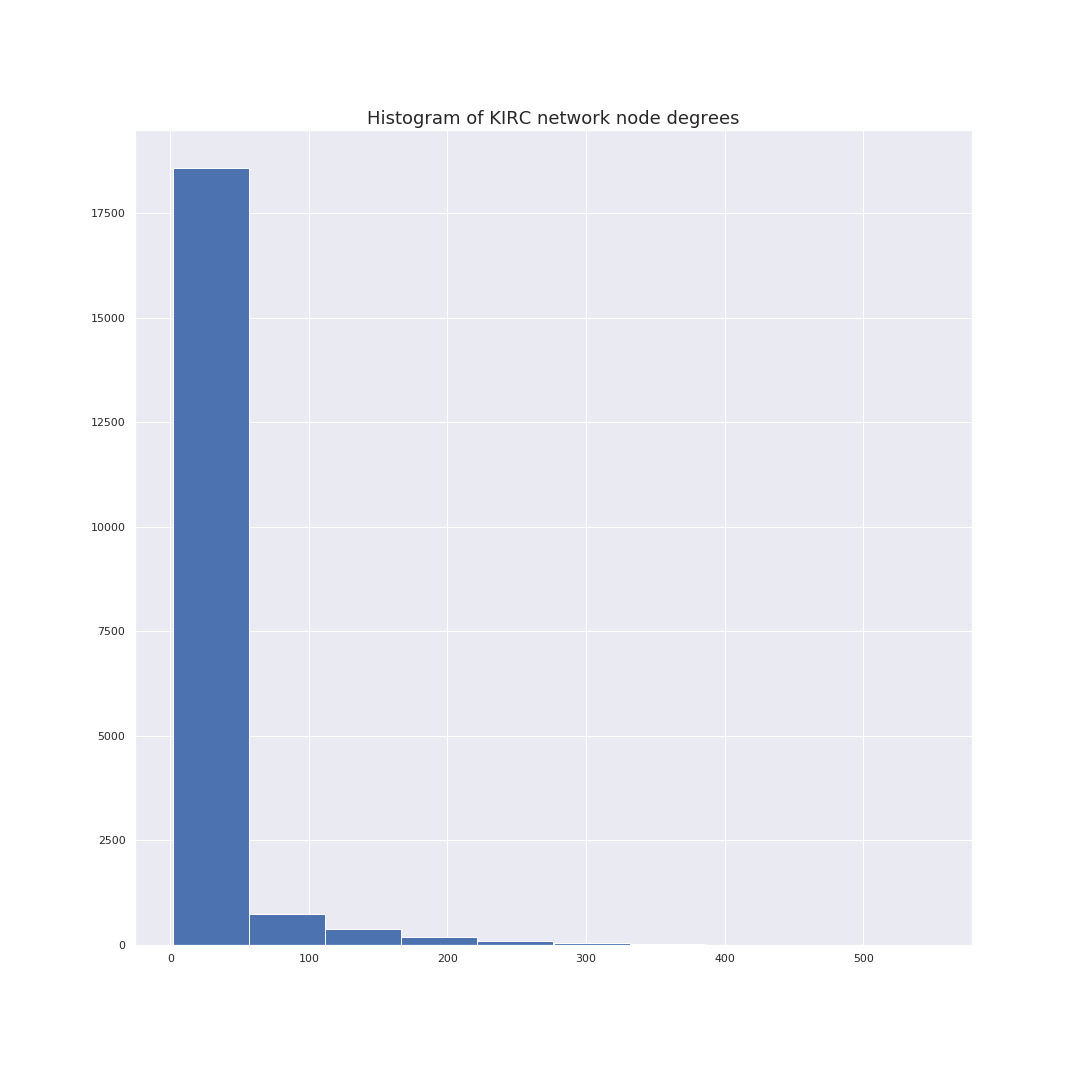}
        \caption{KIRC}
        \label{fig:kircdeg}
    \end{subfigure}
    
    \begin{subfigure}{0.3\linewidth}
        \includegraphics[width=\textwidth]{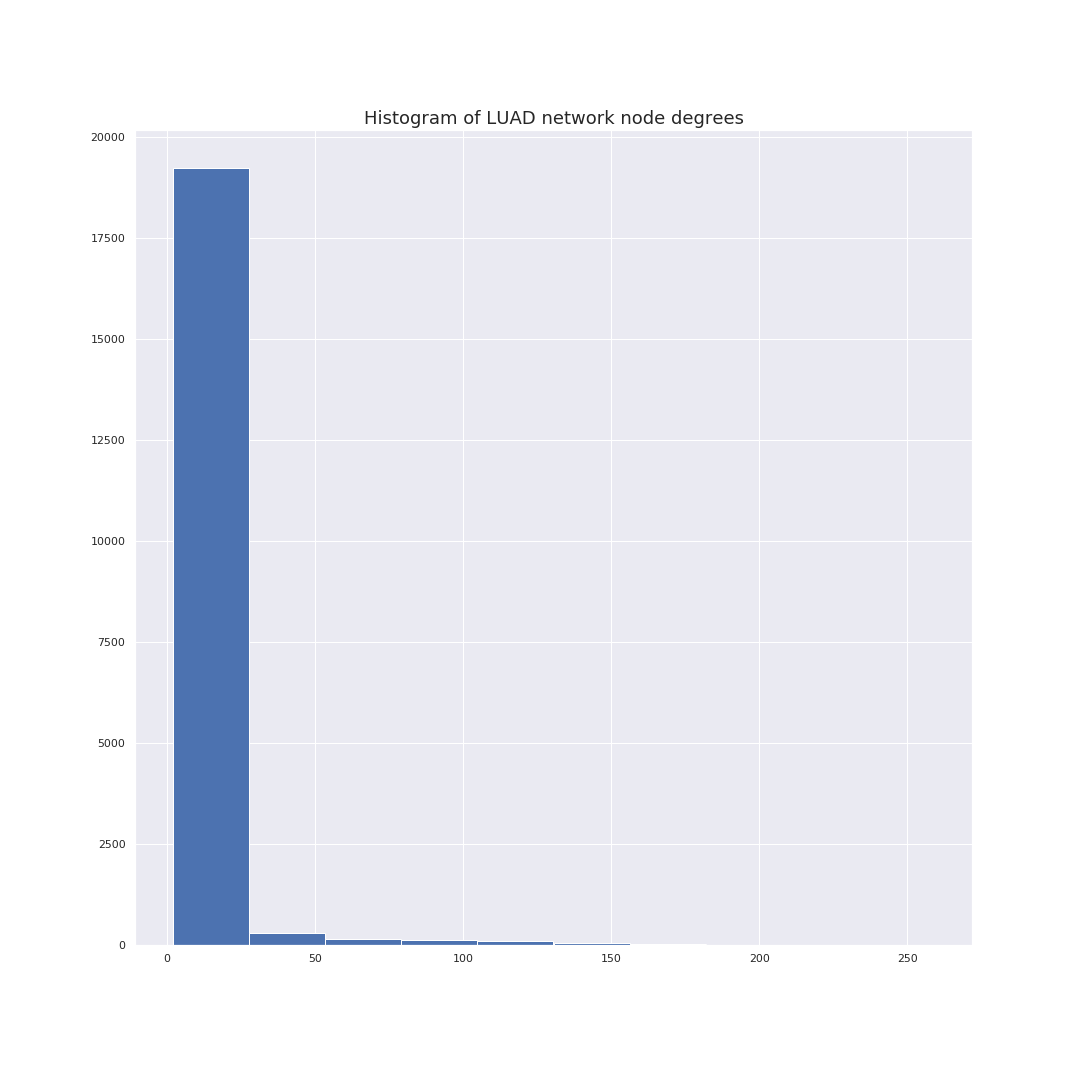}
        \caption{LUAD}
        \label{fig:luaddeg}
    \end{subfigure}
    \begin{subfigure}{0.3\linewidth}
        \includegraphics[width=\textwidth]{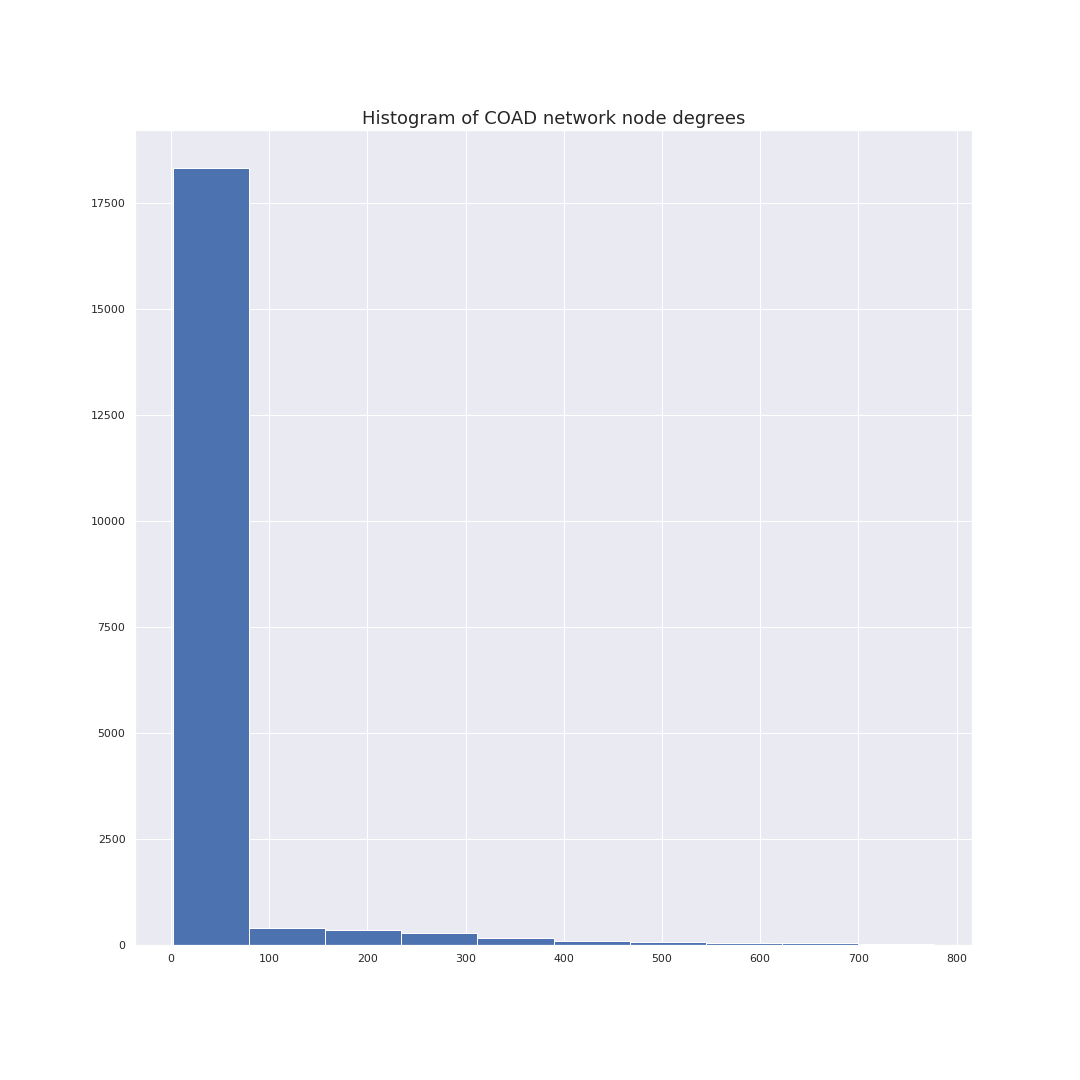}
        \caption{COAD}
        \label{fig:coaddeg}
    \end{subfigure}
    \begin{subfigure}{0.3\linewidth}
        \includegraphics[width=\textwidth]{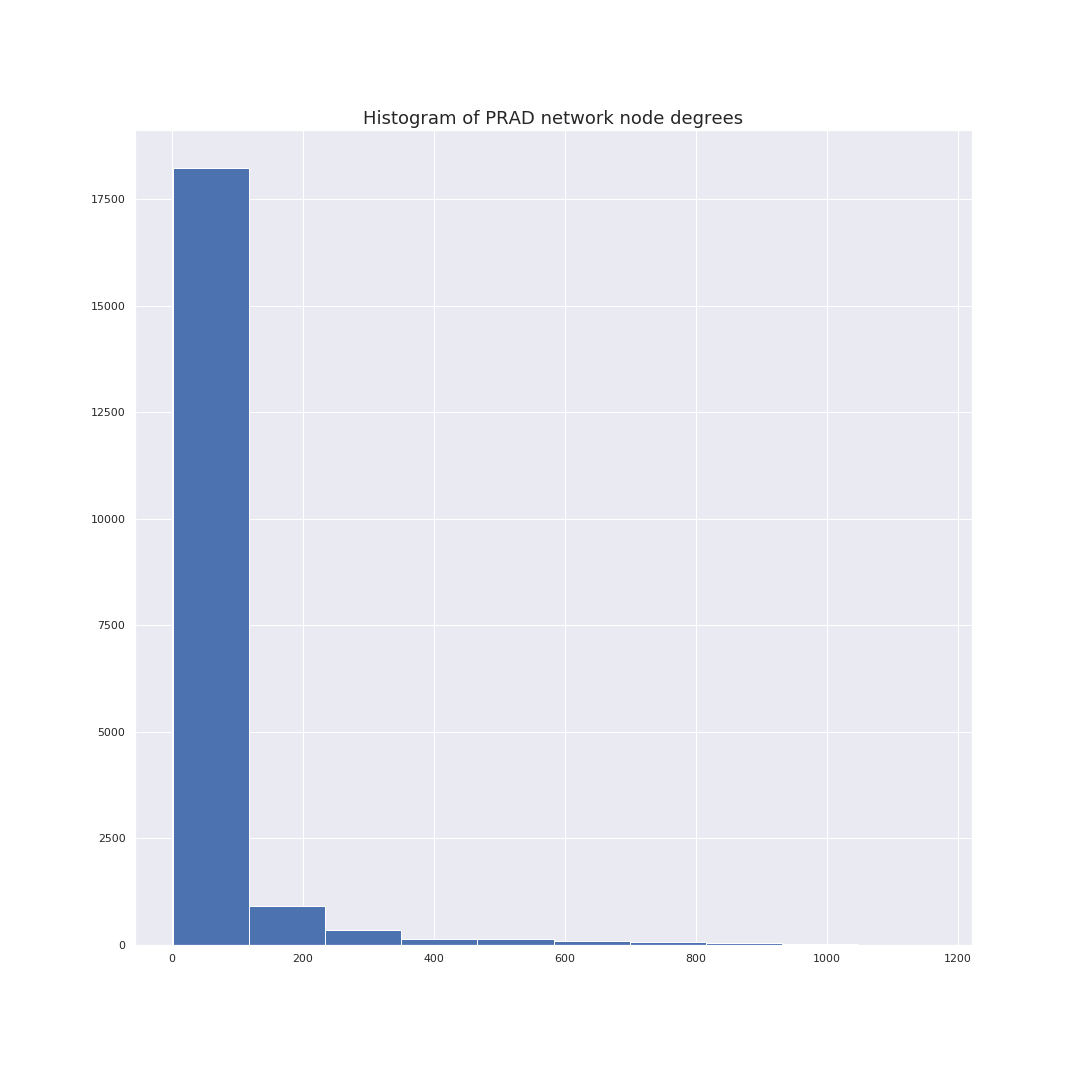}
        \caption{PRAD}
        \label{fig:praddeg}
    \end{subfigure}
    \caption{Degree histogram for each WGCN}
    \label{fig:degreehists}
    \vspace{-5mm}
\end{figure}

Once having the graphs, we first looked at the graph statistics. The basic statistics are summarized in Table~\ref{tab:netstats}. Furthermore, we have plotted the histograms of the degree distributions in Figure~\ref{fig:degreehists}. We can see that that the distributions seem to follow the power law. There seem to be many nodes that have low degrees.

\begin{table}[H]
    \centering
    \begin{tabular}{|c|c|c|c|}
        \hline
        \multicolumn{4}{|c|}{Network descriptions}\\
        \hline
        Cancer Type & \# Nodes & \# Edges  & Avg. Degree\\
        \hline
        BRCA & 20,259 & 43,475 & 4.28 \\
        KIRC & 20,262 & 70,219 & 6.93 \\
        LUAD & 20,251 & 58,963 & 5.82\\
        COAD & 20,227 & 201,408 & 19.91\\
        PRAD & 20,252 & 171,008 & 16.89\\
        \hline
    \end{tabular}
    \caption{Basic statistics of the networks}
    \label{tab:netstats}
\end{table}

After creating those graphs, we computed some centrality measures, such as betweenness centrality, degree centrality and pagerank centrality. The results we got are summarized in tables Table~\ref{pradcent} through Table~\ref{coadcent}, where we can see the gene numbers that ranked higher for each of the centrality measures we computed.

\begin{table}[H]
    \centering
    \begin{tabular}{|c|c|c|c|}
        \hline
        \multicolumn{4}{|c|}{PRAD centralities}\\
        \hline
        Order & Degree & Pagerank & Betweenness\\
        \hline
        1 & 14,974 & 1,671 & 3,068 \\
        2 & 6,799 & 15,985 & 5,177 \\
        3 & 14,643 & 13,761 & 9,525\\
        4 & 5,177 & 19,487 & 19,322\\
        5 & 11,709 & 13,119 & 9,427\\
        \hline
    \end{tabular}
    \caption{Centralities of PRAD WGCN}
    \label{pradcent}
\end{table}

\begin{table}[H]
    \centering
    \begin{tabular}{|c|c|c|c|}
        \hline
        \multicolumn{4}{|c|}{LUAD centralities}\\
        \hline
        Order & Degree & Pagerank & Betweenness\\
        \hline
        1 & 19,819 & 10,462 & 17,124 \\
        2 & 19,582 & 7,749 & 13,269 \\
        3 & 19,196 & 11,394 & 7,502\\
        4 & 18,922& 19,401 & 11,432\\
        5 & 18,918 & 9 & 10,982\\
        \hline
    \end{tabular}
    \caption{Centralities of LUAD WGCN}
    \label{luadcent}
\end{table}

\begin{table}[H]
    \centering
    \begin{tabular}{|c|c|c|c|}
        \hline
        \multicolumn{4}{|c|}{BRCA centralities}\\
        \hline
        Order & Degree & Pagerank & Betweenness\\
        \hline
        1 & 15,512 & 14,974 & 19,862 \\
        2 & 14,376 & 17,430 & 4,749 \\
        3 & 3,356 & 16,274 & 14,974\\
        4 & 8,355& 19,847 & 20,355\\
        5 & 1,139 & 715 & 1,511\\
        \hline
    \end{tabular}
    \caption{Centralities of BRCA WGCN}
    \label{brcacent}
\end{table}

\begin{table}[H]
    \centering
    \begin{tabular}{|c|c|c|c|}
        \hline
        \multicolumn{4}{|c|}{KIRC centralities}\\
        \hline
        Order & Degree & Pagerank & Betweenness\\
        \hline
        1 & 6,799 & 1,363 & 15,147 \\
        2 & 2,111 & 18,173 & 19,309 \\
        3 & 6,022 & 2,124 & 17,805\\
        4 & 3,267& 1,298 & 5,330\\
        5 & 17,791 & 3,913 & 13,650\\
        \hline
    \end{tabular}
    \caption{Centralities of KIRC WGCN}
    \label{kirccent}
\end{table}

\begin{table}[H]
    \centering
    \begin{tabular}{|c|c|c|c|}
        \hline
        \multicolumn{4}{|c|}{COAD centralities}\\
        \hline
        Order & Degree & Pagerank & Betweenness\\
        \hline
        1 & 19,375 & 12,509 & 1,213 \\
        2 & 6,259 & 12,402 & 16,556 \\
        3 & 18,822 & 5,280 & 16,463\\
        4 & 3,997& 4 & 5,198\\
        5 & 19,862 & 15,139 & 713\\
        \hline
    \end{tabular}
    \caption{Centralities of COAD WGCN}
    \label{coadcent}
\end{table}

From this analysis, we can see what the most "important" genes are for each cancer type, based on their centralities. The genes with the highest centralities will be the most prominent in patients with the respective type of cancer, producing an outsize effect on the overall gene expression. We could then try to map each of the gene numbers to the actual gene name by ordering the gene sequence and finding the gene corresponding to each index number. From there we could research the function of that gene. We expect the function of the genes with the highest centrality in each cancer type to be somehow related to that organ in the body. 

Furthermore, it is interesting to see that it is not the case that the top 5 genes are the same in each centrality measures. This happens because each centrality measure is computed differently and will lead to different results. Moreover, there is a very large amount of genes many of whom have very similar values for their centrality scores which means that even though one gene ranking highly in one centrality measure could have a very high centrality score in a different measure, it might still not make it in the ``top 5".

\begin{figure}[H]
    \centering
    \includegraphics[trim={0 0 0 3cm},clip,width = \linewidth]{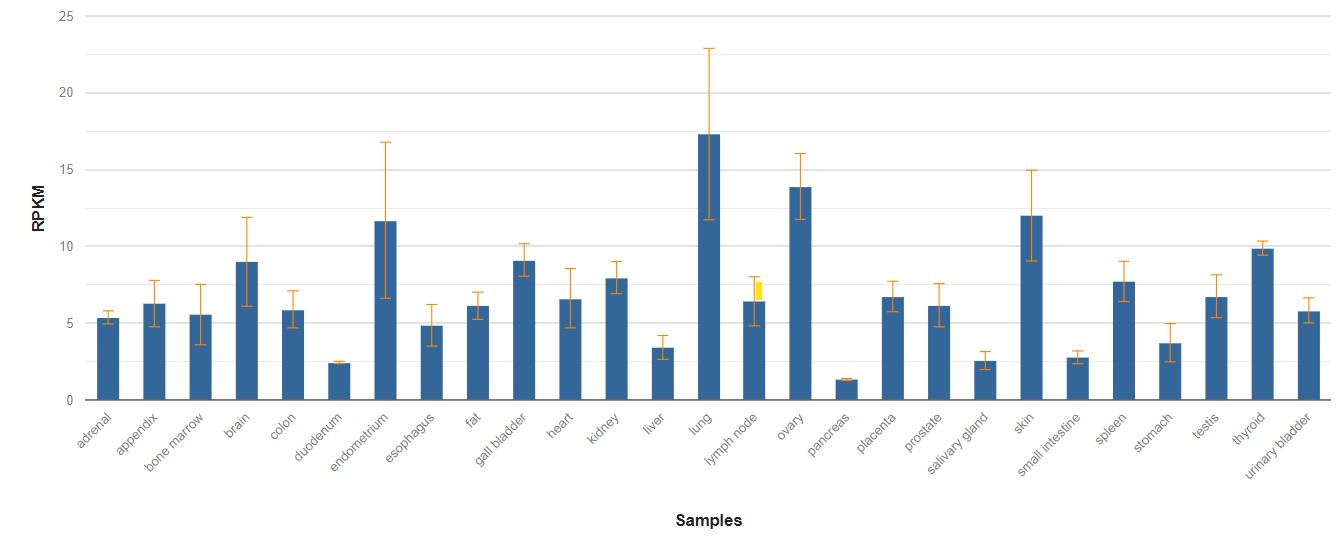}
    \caption{Gene expression comparison}
    \label{fig:lungexp}
\end{figure}

Using the list of the gene names and mapping that to the index given to us, we can find what gene name each gene index number corresponds to. Then we can search on the National Center for Biotechnology Information (NCBI), we can find what exactly each gene does and where it is most expressed. For example, the gene with the highest pagerank centrality in LUAD (lung cancer) is gene \#10,462 which corresponds to gene "MACF1 23499". NCBI tells us that this gene "encodes a large protein which is a member of a family of proteins that form bridges between different cytoskeletal elements". Furthermore when we see in general this gene is mostly expressed in lung tissue, as shown in figure~\ref{fig:lungexp}. Furthermore gene \#17,124 (highest betweenness in lung cancer) or "SPEN 23013" is a transcriptional repressor, which would make sense to have high centrality that regulates multiple genes expression that related to cancer.

\begin{figure}[H]
    \centering
    \includegraphics[trim={0 0 0 4cm},clip, width = \linewidth]{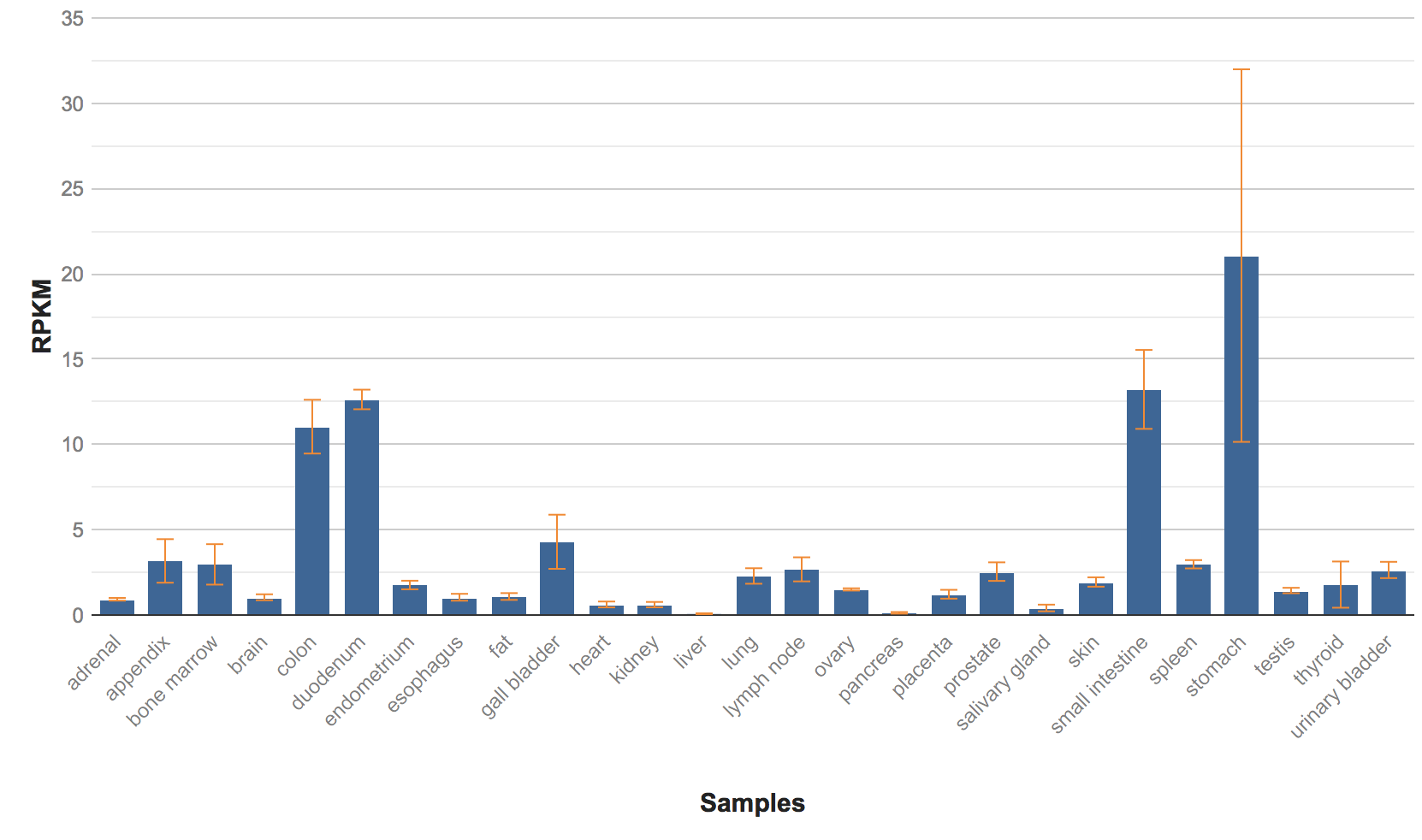}
    \caption{Gene expression comparison}
    \label{fig:colonexp}
\end{figure}

Additionally, the gene with the highest degree centrality in colon cancer (VILL 50853), shows most expression in the stomach area, intestines and colon, as shown in Figure~\ref{fig:colonexp}.


\section{Clusters Comparison}

\subsection{Identifying the Critical Subset of Genes}
We fit a multinomial logistic regression model to classify our data, estimating coefficients for each gene. Analyzing these coefficients, we can determine whether that gene is a statistically significant determinant of a particular cancer type. Multinomial logistic regression is the generalization of logistic regression to multiple categories. Since we do not have normal samples in our dataset, we use BRCA samples to indicate a baseline, since it is the plurality of our samples. Fitting this model, we get both coefficients and standard errors, and each number corresponding to a model equation.  For example, the first row (COAD) for the gene\# is regressed to the equation.

$$ ln(\frac{P(cancer type=COAD)}{P(cancer type=BRCA)})=\beta_{0}+\beta\cdot gene\# $$

The way to interpret this regression is that $\beta$ means one unit increase in gene\# is associated with the decrease of probability in being COAD instead of BRCA in the amount of ~$\beta$. To be more specific, the ratio of the probability of choose one outcome category over the probability of choose the baseline category is the right-hand side linear equation exponentiated. 

$$ \frac{P(cancer type=COAD)}{P(cancer type=BRCA)}=a\cdot e^{\beta\cdot gene \#} $$

Thus, $\beta$ are relative risk ratios for a unit change of predictor variable. Since we have 801 samples data, this should provide a reasonably accurate estimate through regression. We also got standard deviation from the regression processes (for the coefficient)  P-values were calculated according to t-tests $H_{0}:\beta=0$ vs.$H_{A}:\beta \neq 0$. 

After we got the P-values for all cancer types based on breast cancer over 20532 genes. We chose the significance level to be 0.005 and delete the genes with P-values below this threshold in all 4 cancer types. The further analysis is based on this small dataset. The $-\log(p)$ vs. gene\# were plotted (known as Manhattan plots) for each type of cancer. 

\begin{figure}[H]
    \centering
    \includegraphics[width = \linewidth]{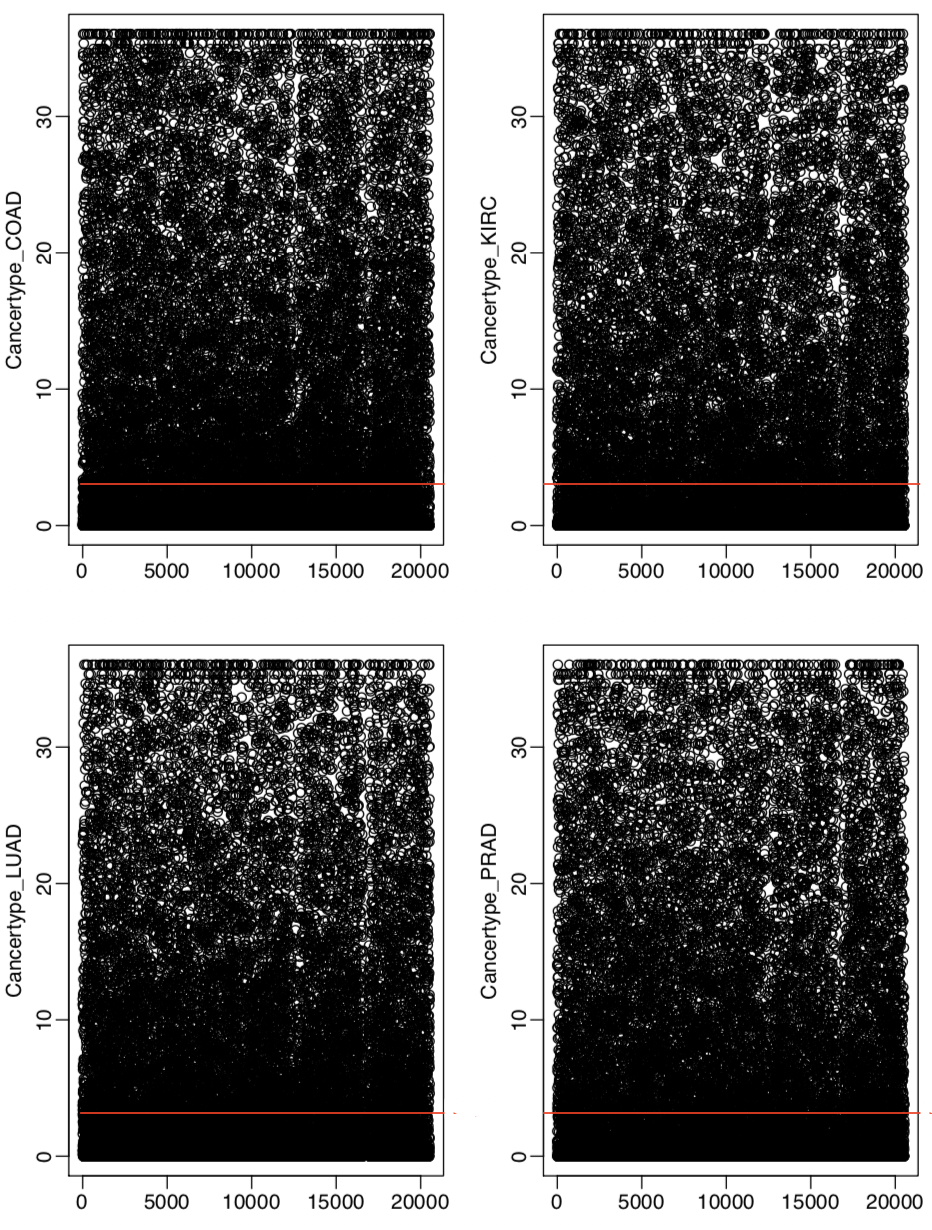}
    \caption{–Log(p) vs. Gene\# of 4 cancer types COAD,KIRC, LUAD, PRAD. Red line (–log(p)=2.3) is the threshold. Genes below this threshold were removed.}
    \label{fig:manhattan plots}
\end{figure}

By choosing a threshold of p=0.005, –log(p)=2.3. We removed data below the red line in the plot. However, according to the plots there are still lots of genes to be analyzed. 
Interestingly, there are two narrow blank spaces shown in all the plots and those parts may suggest that P value are all large, and the cancer types has no relationship with those genes. Thus, those genes can either be genes related to this cancer (and are similar regulated in all cancer types) or they can be genes unrelated to this cancer (similar expressed for all people). 
This method helped to reduce the set to 1075 genes, which we used for the following analysis.

\subsection{Clustering by Gene Expression Levels}
Now with the smaller dataset, we want to analyze the expression level of genes among different cancer types, specifically we focused on LUAD and PRAD. They are chosen since they have similar sample size. Before any further analysis, the gene expressions were normalized according to the average and standard deviation of that specific gene expression in all cancer types. 

The expression levels distributed according to the histograms in LUAD, PRAD.

\begin{figure}[H]
    \centering
    \includegraphics[width = \linewidth]{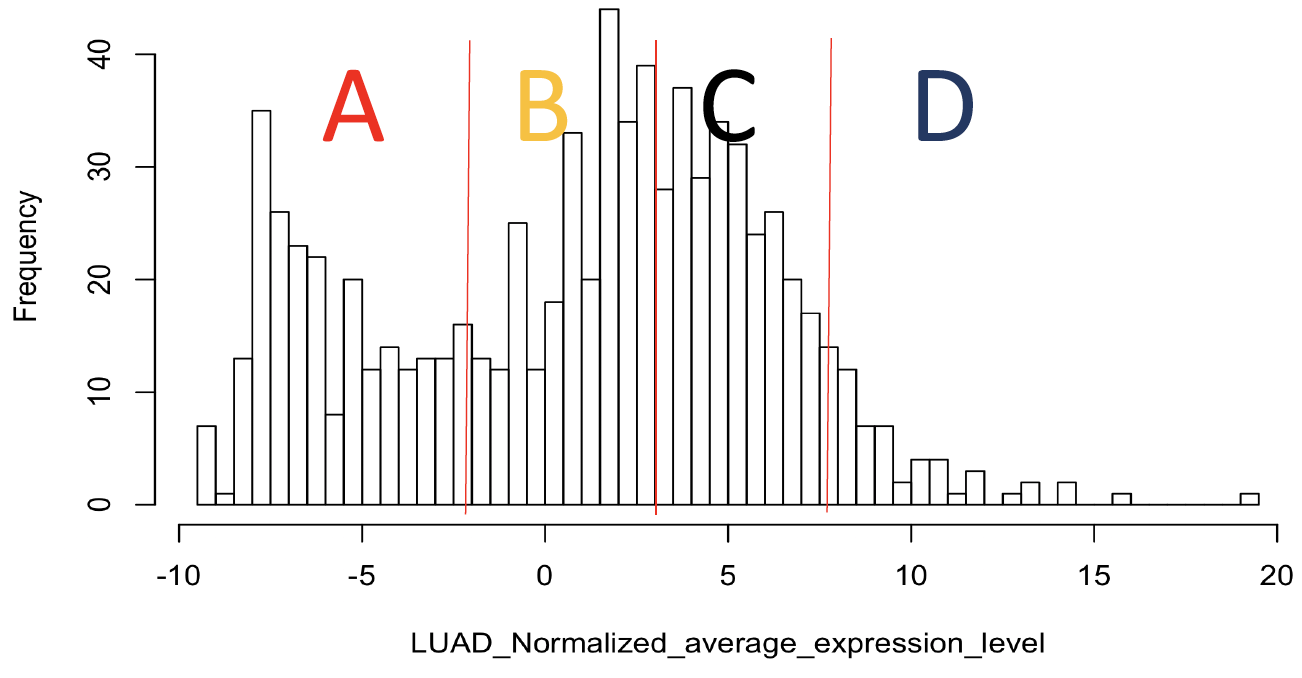}
    \caption{Histogram of genes expression levels in LUAD. A down-, B normal-, C over-, D highly expressed.}
    \label{fig:expressionlevelplot1}
\end{figure}

\begin{figure}[H]
    \centering
    \includegraphics[width = \linewidth]{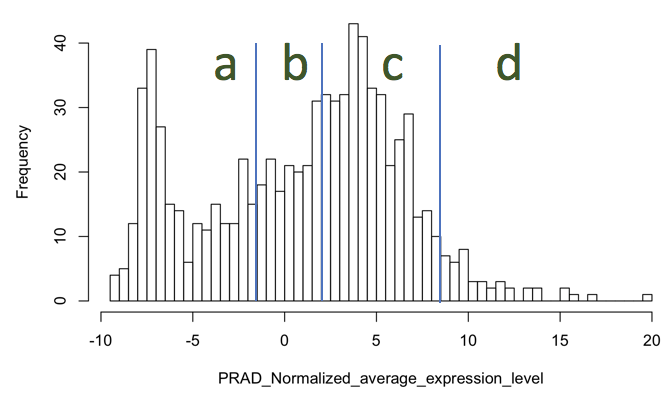}
    \caption{Histogram of genes expression levels in PRAD. a down-, b normal-, c over-, d highly expressed.}
    \label{fig:expressionlevelplot2}
\end{figure}

According to the graph, we think the expression could be grouped into 5 groups, group A: (-10,-2), group B: (-2,2), group C: (2, 8), group D: more than 8, with other NA values to be group 0. This is consistent with the down-regulated genes (compared with other cancer types), normal-expressed genes, slightly over-expressed genes, highly over-expressed genes. (This could be changed according to the tissue type to make it more biological meaningful.) 

\subsection{Conserved of Gene Expression Levels in LUAD and PRAD}
According to the groups, an adjacency matrix is generated with 1 at $a_{ij}$ if both gene $i$ and gene $j$ are in same group, otherwise 0. This is then used to generate network. We expected to see all the genes in same group would be fully connected with each other, and a function were used to separate the connected components apart. The networks of LUAD and PRAD were shown as following. 

\begin{figure}[H]
    \centering
    \includegraphics[width = \linewidth]{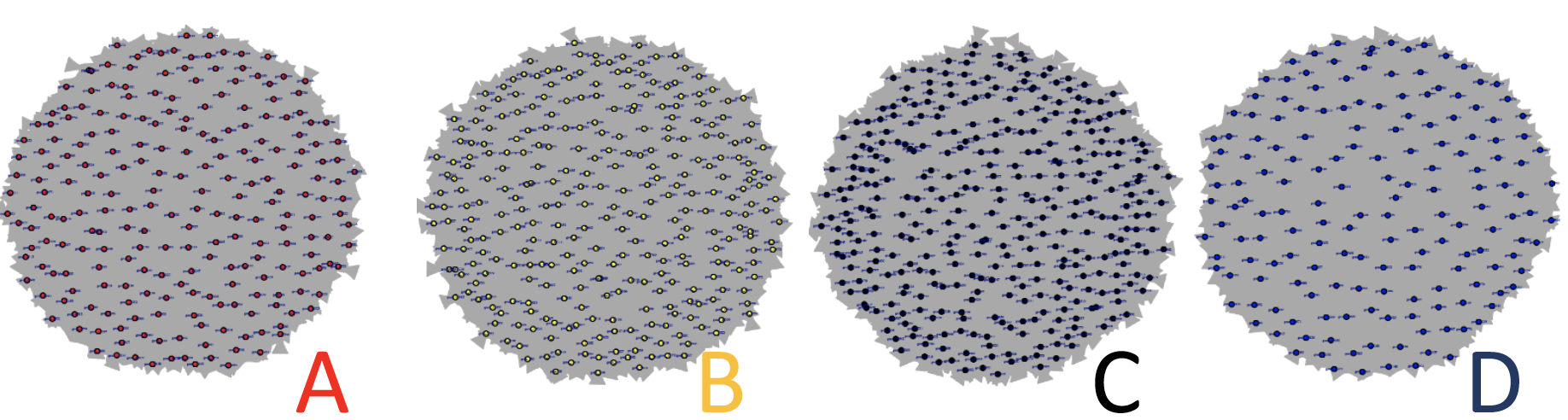}
    \caption{LUAD Network with gene in same label(A B C D see Histogram) fully connected.}
    \label{fig:expressionlevelnetwork1}
\end{figure}

\begin{figure}[H]
    \centering
   \includegraphics[width = \linewidth]{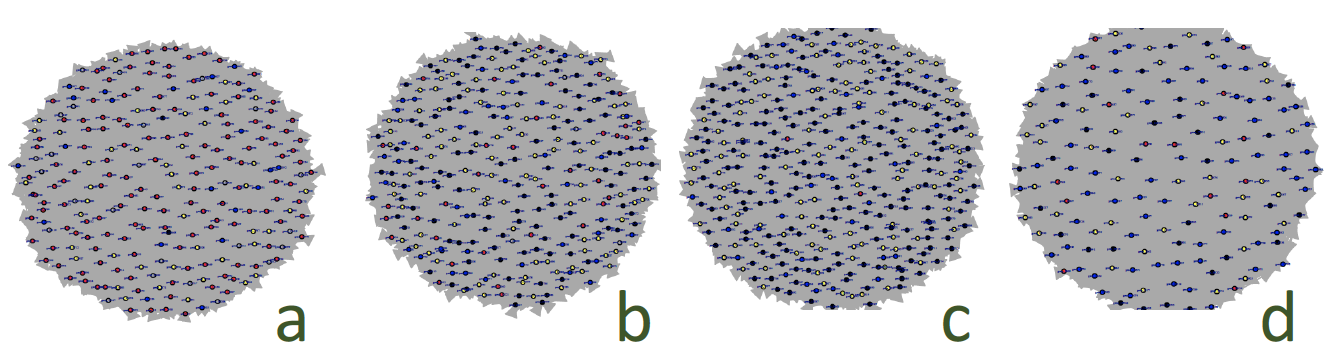}
    \caption{PRAD Network with gene in same label(a b c d see Histogram) fully connected with color of the nodes consistent with LUAD labels}
    \label{fig:expressionlevelnetwork2}
\end{figure}

The nodes are colored according to the expression level in LUAD cancer: group A as red, B as yellow, C as black and D as blue (consistent with their labels in LUAD histogram). The graphs suggest that in general, the genes expressed highly in LUAD are distributed evenly in PRAD, and especially group B and group C are mixed evenly. Probably because the threshold choose is not significant. However, there are some preservation of the expression patterns. 

\section{Discussion and Further Work}

\begin{table}[H]
    \centering
    \begin{tabular}{|c|c|c|c|c|}
        \hline
        Rank & Type & New Cases & \% & Deadliest  \\
        \hline 
        1 & Lung  & 2,093,876 & 12.3 & 1 \\
        2 & Breast  & 2,088,849 & 12.3 & 3 \\
        3 & Colorectal  & 1,800,977 & 10.6 & 2 \\
        4 & Prostate  & 1,276,106 & 7.5 & 5 \\
        5 & Stomach  & 1,033,701 & 6.1 & n/a\\
        \hline
    \end{tabular}
    \caption{Global cancer incidence \cite{bray2018global} where \% refers to the percent of new cases of cancer diagnosed in the US in 2018 and deadliest refers to the ranking for that cancer in causing the most deaths in 2018}
    \label{tab:incidence}
\end{table}

Cancer is one of the most significant public health challenges, particularly in the developed world. In this project, we examined 4 of the top 5 (and 5 of the top 15) cancer types in terms of new cases diagnosed in 2018, evidenced by Table \ref{tab:incidence}. In addition to being prevalent, the cancer types studied here correspond to 4 of the top 5 cancer types contributing to deaths in America. The prevalence of datasets and computational tools has revolutionized nearly all fields of science, particularly biology. Transferring successful models from the statistical modeling literature to this dataset has allowed us to validate existing scientific conclusions and identify areas which warrant further study. We were pleased to find that the cancer types can be clustered into groups using out-of-the-box approaches for dimensionality reduction. Since each cancer type is different in many ways, it is reassuring to see those differences reflected in our statistical approach. Other portions of our report highlight areas that could be worth exploring further from the biomedical perspective. For example, in section 2.3, it was seen that the 4 most prevalent and deadly cancer types appeared clustered more closely together than to KIRC. It would be interesting to explore how this matches the intuition of oncologists, who might have a sense of which cancer types are more similar to each other. Through our network approaches we were able to identify genes of interest. We are optimistic that these centrality measure of our network correspond to biological insight and that these network approaches can serve as a spotlight to help guide researchers to study potentially high impact areas of the genome in a more efficient manner.

\section{Conclusion}

In this report we were able to summarize the similarities and differences of 801 samples of 5 cancer types from a dataset generated by UC Irvine. After performing an exploratory analysis, we were surprised to see that the gene expression profiles could be easily clustered. This motivated further analysis to characterize interactions between patients and genes that were indicative of biological differences between the cancer types. To characterize these relationships we constructed networks: one that represented the relationships between patients and another that aimed to characterize the relationships between genes. Using standard network analysis measures (such as centrality statistics) we highlighted genes that appear to be highly influential for each cancer type, such as MACF1 23499 for LUAD and VILL 50853 for COAD. Both these genes appear to plausible genes involved with these cancer types, validating elements of our approach. These networks approaches and expression analyses applied gene expression data aim to motivate for future work to understand the biological implications of standard statistical measures in gene expression profiles.


\nocite{*}
\bibliography{aipsamp}

\begin{thebibliography}{10}%
\makeatletter
\providecommand \@ifxundefined [1]{%
 \@ifx{#1\undefined}
}%
\providecommand \@ifnum [1]{%
 \ifnum #1\expandafter \@firstoftwo
 \else \expandafter \@secondoftwo
 \fi
}%
\providecommand \@ifx [1]{%
 \ifx #1\expandafter \@firstoftwo
 \else \expandafter \@secondoftwo
 \fi
}%
\providecommand \natexlab [1]{#1}%
\providecommand \enquote  [1]{``#1''}%
\providecommand \bibnamefont  [1]{#1}%
\providecommand \bibfnamefont [1]{#1}%
\providecommand \citenamefont [1]{#1}%
\providecommand \href@noop [0]{\@secondoftwo}%
\providecommand \href [0]{\begingroup \@sanitize@url \@href}%
\providecommand \@href[1]{\@@startlink{#1}\@@href}%
\providecommand \@@href[1]{\endgroup#1\@@endlink}%
\providecommand \@sanitize@url [0]{\catcode `\\12\catcode `\$12\catcode
  `\&12\catcode `\#12\catcode `\^12\catcode `\_12\catcode `\%12\relax}%
\providecommand \@@startlink[1]{}%
\providecommand \@@endlink[0]{}%
\providecommand \url  [0]{\begingroup\@sanitize@url \@url }%
\providecommand \@url [1]{\endgroup\@href {#1}{\urlprefix }}%
\providecommand \urlprefix  [0]{URL }%
\providecommand \Eprint [0]{\href }%
\providecommand \doibase [0]{http://dx.doi.org/}%
\providecommand \selectlanguage [0]{\@gobble}%
\providecommand \bibinfo  [0]{\@secondoftwo}%
\providecommand \bibfield  [0]{\@secondoftwo}%
\providecommand \translation [1]{[#1]}%
\providecommand \BibitemOpen [0]{}%
\providecommand \bibitemStop [0]{}%
\providecommand \bibitemNoStop [0]{.\EOS\space}%
\providecommand \EOS [0]{\spacefactor3000\relax}%
\providecommand \BibitemShut  [1]{\csname bibitem#1\endcsname}%
\let\auto@bib@innerbib\@empty
\bibitem [{\citenamefont {Khan}\ \emph {et~al.}(2001)\citenamefont {Khan},
  \citenamefont {Wei}, \citenamefont {Ringner}, \citenamefont {Saal},
  \citenamefont {Ladanyi}, \citenamefont {Westermann}, \citenamefont
  {Berthold}, \citenamefont {Schwab}, \citenamefont {Antonescu}, \citenamefont
  {Peterson} \emph {et~al.}}]{khan2001classification}%
  \BibitemOpen
  \bibfield  {author} {\bibinfo {author} {\bibfnamefont {J.}~\bibnamefont
  {Khan}}, \bibinfo {author} {\bibfnamefont {J.~S.}\ \bibnamefont {Wei}},
  \bibinfo {author} {\bibfnamefont {M.}~\bibnamefont {Ringner}}, \bibinfo
  {author} {\bibfnamefont {L.~H.}\ \bibnamefont {Saal}}, \bibinfo {author}
  {\bibfnamefont {M.}~\bibnamefont {Ladanyi}}, \bibinfo {author} {\bibfnamefont
  {F.}~\bibnamefont {Westermann}}, \bibinfo {author} {\bibfnamefont
  {F.}~\bibnamefont {Berthold}}, \bibinfo {author} {\bibfnamefont
  {M.}~\bibnamefont {Schwab}}, \bibinfo {author} {\bibfnamefont {C.~R.}\
  \bibnamefont {Antonescu}}, \bibinfo {author} {\bibfnamefont {C.}~\bibnamefont
  {Peterson}},  \emph {et~al.},\ }\href@noop {} {\bibfield  {journal} {\bibinfo
   {journal} {Nature medicine}\ }\textbf {\bibinfo {volume} {7}},\ \bibinfo
  {pages} {673} (\bibinfo {year} {2001})}\BibitemShut {NoStop}%
\bibitem [{\citenamefont {Lee}\ and\ \citenamefont
  {Lee}(2003)}]{lee2003classification}%
  \BibitemOpen
  \bibfield  {author} {\bibinfo {author} {\bibfnamefont {Y.}~\bibnamefont
  {Lee}}\ and\ \bibinfo {author} {\bibfnamefont {C.-K.}\ \bibnamefont {Lee}},\
  }\href@noop {} {\bibfield  {journal} {\bibinfo  {journal} {Bioinformatics}\
  }\textbf {\bibinfo {volume} {19}},\ \bibinfo {pages} {1132} (\bibinfo {year}
  {2003})}\BibitemShut {NoStop}%
\bibitem [{\citenamefont {Tibshirani}\ \emph {et~al.}(2002)\citenamefont
  {Tibshirani}, \citenamefont {Hastie}, \citenamefont {Narasimhan},\ and\
  \citenamefont {Chu}}]{tibshirani2002diagnosis}%
  \BibitemOpen
  \bibfield  {author} {\bibinfo {author} {\bibfnamefont {R.}~\bibnamefont
  {Tibshirani}}, \bibinfo {author} {\bibfnamefont {T.}~\bibnamefont {Hastie}},
  \bibinfo {author} {\bibfnamefont {B.}~\bibnamefont {Narasimhan}}, \ and\
  \bibinfo {author} {\bibfnamefont {G.}~\bibnamefont {Chu}},\ }\href@noop {}
  {\bibfield  {journal} {\bibinfo  {journal} {Proceedings of the National
  Academy of Sciences}\ }\textbf {\bibinfo {volume} {99}},\ \bibinfo {pages}
  {6567} (\bibinfo {year} {2002})}\BibitemShut {NoStop}%
\bibitem [{\citenamefont {Dubey}, \citenamefont {Gupta},\ and\ \citenamefont
  {Jain}(2015)}]{dubey2015breast}%
  \BibitemOpen
  \bibfield  {author} {\bibinfo {author} {\bibfnamefont {A.~K.}\ \bibnamefont
  {Dubey}}, \bibinfo {author} {\bibfnamefont {U.}~\bibnamefont {Gupta}}, \ and\
  \bibinfo {author} {\bibfnamefont {S.}~\bibnamefont {Jain}},\ }\href@noop {}
  {\bibfield  {journal} {\bibinfo  {journal} {Asian Pac J Cancer Prev}\
  }\textbf {\bibinfo {volume} {16}},\ \bibinfo {pages} {4237} (\bibinfo {year}
  {2015})}\BibitemShut {NoStop}%
\bibitem [{\citenamefont {Botia}\ \emph {et~al.}(2018)\citenamefont {Botia},
  \citenamefont {Guelfi}, \citenamefont {Zhang}, \citenamefont {D'Sa},
  \citenamefont {Reinolds}, \citenamefont {Onah}, \citenamefont {McDonagh},
  \citenamefont {Rueda-Martin}, \citenamefont {Tucci}, \citenamefont {Rendon}
  \emph {et~al.}}]{botia2018g2p}%
  \BibitemOpen
  \bibfield  {author} {\bibinfo {author} {\bibfnamefont {J.~A.}\ \bibnamefont
  {Botia}}, \bibinfo {author} {\bibfnamefont {S.}~\bibnamefont {Guelfi}},
  \bibinfo {author} {\bibfnamefont {D.}~\bibnamefont {Zhang}}, \bibinfo
  {author} {\bibfnamefont {K.}~\bibnamefont {D'Sa}}, \bibinfo {author}
  {\bibfnamefont {R.}~\bibnamefont {Reinolds}}, \bibinfo {author}
  {\bibfnamefont {D.}~\bibnamefont {Onah}}, \bibinfo {author} {\bibfnamefont
  {E.~M.}\ \bibnamefont {McDonagh}}, \bibinfo {author} {\bibfnamefont
  {A.}~\bibnamefont {Rueda-Martin}}, \bibinfo {author} {\bibfnamefont
  {A.}~\bibnamefont {Tucci}}, \bibinfo {author} {\bibfnamefont
  {A.}~\bibnamefont {Rendon}},  \emph {et~al.},\ }\href@noop {} {\bibfield
  {journal} {\bibinfo  {journal} {bioRxiv}\ ,\ \bibinfo {pages} {288845}}
  (\bibinfo {year} {2018})}\BibitemShut {NoStop}%
\bibitem [{\citenamefont {Yang}\ \emph {et~al.}(2014)\citenamefont {Yang},
  \citenamefont {Han}, \citenamefont {Yuan}, \citenamefont {Li}, \citenamefont
  {Hei},\ and\ \citenamefont {Liang}}]{yang2014gene}%
  \BibitemOpen
  \bibfield  {author} {\bibinfo {author} {\bibfnamefont {Y.}~\bibnamefont
  {Yang}}, \bibinfo {author} {\bibfnamefont {L.}~\bibnamefont {Han}}, \bibinfo
  {author} {\bibfnamefont {Y.}~\bibnamefont {Yuan}}, \bibinfo {author}
  {\bibfnamefont {J.}~\bibnamefont {Li}}, \bibinfo {author} {\bibfnamefont
  {N.}~\bibnamefont {Hei}}, \ and\ \bibinfo {author} {\bibfnamefont
  {H.}~\bibnamefont {Liang}},\ }\href@noop {} {\bibfield  {journal} {\bibinfo
  {journal} {Nature communications}\ }\textbf {\bibinfo {volume} {5}},\
  \bibinfo {pages} {3231} (\bibinfo {year} {2014})}\BibitemShut {NoStop}%
\bibitem [{\citenamefont {Maaten}\ and\ \citenamefont
  {Hinton}(2008)}]{maaten2008visualizing}%
  \BibitemOpen
  \bibfield  {author} {\bibinfo {author} {\bibfnamefont {L.~v.~d.}\
  \bibnamefont {Maaten}}\ and\ \bibinfo {author} {\bibfnamefont
  {G.}~\bibnamefont {Hinton}},\ }\href@noop {} {\bibfield  {journal} {\bibinfo
  {journal} {Journal of machine learning research}\ }\textbf {\bibinfo {volume}
  {9}},\ \bibinfo {pages} {2579} (\bibinfo {year} {2008})}\BibitemShut
  {NoStop}%
\bibitem [{\citenamefont {Fuller}\ \emph {et~al.}(2007)\citenamefont {Fuller},
  \citenamefont {Ghazalpour}, \citenamefont {Aten}, \citenamefont {Drake},
  \citenamefont {Lusis},\ and\ \citenamefont {Horvath}}]{fuller2007weighted}%
  \BibitemOpen
  \bibfield  {author} {\bibinfo {author} {\bibfnamefont {T.~F.}\ \bibnamefont
  {Fuller}}, \bibinfo {author} {\bibfnamefont {A.}~\bibnamefont {Ghazalpour}},
  \bibinfo {author} {\bibfnamefont {J.~E.}\ \bibnamefont {Aten}}, \bibinfo
  {author} {\bibfnamefont {T.~A.}\ \bibnamefont {Drake}}, \bibinfo {author}
  {\bibfnamefont {A.~J.}\ \bibnamefont {Lusis}}, \ and\ \bibinfo {author}
  {\bibfnamefont {S.}~\bibnamefont {Horvath}},\ }\href@noop {} {\bibfield
  {journal} {\bibinfo  {journal} {Mammalian Genome}\ }\textbf {\bibinfo
  {volume} {18}},\ \bibinfo {pages} {463} (\bibinfo {year} {2007})}\BibitemShut
  {NoStop}%
\bibitem [{\citenamefont {Langfelder}\ and\ \citenamefont
  {Horvath}(2008)}]{langfelder2008wgcna}%
  \BibitemOpen
  \bibfield  {author} {\bibinfo {author} {\bibfnamefont {P.}~\bibnamefont
  {Langfelder}}\ and\ \bibinfo {author} {\bibfnamefont {S.}~\bibnamefont
  {Horvath}},\ }\href@noop {} {\bibfield  {journal} {\bibinfo  {journal} {BMC
  bioinformatics}\ }\textbf {\bibinfo {volume} {9}},\ \bibinfo {pages} {559}
  (\bibinfo {year} {2008})}\BibitemShut {NoStop}%
\bibitem [{\citenamefont {Bray}\ \emph {et~al.}(2018)\citenamefont {Bray},
  \citenamefont {Ferlay}, \citenamefont {Soerjomataram}, \citenamefont
  {Siegel}, \citenamefont {Torre},\ and\ \citenamefont
  {Jemal}}]{bray2018global}%
  \BibitemOpen
  \bibfield  {author} {\bibinfo {author} {\bibfnamefont {F.}~\bibnamefont
  {Bray}}, \bibinfo {author} {\bibfnamefont {J.}~\bibnamefont {Ferlay}},
  \bibinfo {author} {\bibfnamefont {I.}~\bibnamefont {Soerjomataram}}, \bibinfo
  {author} {\bibfnamefont {R.~L.}\ \bibnamefont {Siegel}}, \bibinfo {author}
  {\bibfnamefont {L.~A.}\ \bibnamefont {Torre}}, \ and\ \bibinfo {author}
  {\bibfnamefont {A.}~\bibnamefont {Jemal}},\ }\href@noop {} {\bibfield
  {journal} {\bibinfo  {journal} {CA: a cancer journal for clinicians}\
  }\textbf {\bibinfo {volume} {68}},\ \bibinfo {pages} {394} (\bibinfo {year}
  {2018})}\BibitemShut {NoStop}%
\end{thebibliography}%

\end{document}